\documentclass[12pt,preprint]{aastex}

\shorttitle{White Dwarfs}
\shortauthors{Hurley \& Shara}

\slugcomment{To appear in the Astrophysical Journal}

\begin{document}

\title{White Dwarf Sequences in Dense Star Clusters}

\author{Jarrod R. Hurley\altaffilmark{1} and Michael M. Shara}
\affil{Department of Astrophysics, 
       American Museum of Natural History, \\ 
       Central Park West at 79th Street, 
       New York, NY 10024}
\altaffiltext{1}{Hubble Fellow}
\email{jhurley@amnh.org, mshara@amnh.org}

\begin{abstract}
We use the results of realistic $N$-body simulations to investigate 
the appearance of the white dwarf population in dense star clusters. 
We show that the presence of a substantial binary population in a 
star cluster, and the interaction of this population with the 
cluster environment, has serious consequences for the morphology of 
the observed white dwarf sequence and the derived white dwarf cooling 
age of the cluster. 
We find that over time the dynamical evolution of the cluster 
-- mass-segregation, stellar interactions and tidal stripping -- 
hampers the use of white dwarfs as tracers of the initial mass function, 
and also leads to a significant enhancement of the white dwarf 
mass fraction. 
Future observations of star clusters should be conducted slightly 
interior to the half-mass radius of the cluster in order to best 
obtain information about the cluster age and initial mass function 
from the white dwarf luminosity function. 
The evolution of binary stars and the cluster environment must 
necessarily be accounted for when studying the white dwarf 
populations of dynamically evolved star clusters. 
\end{abstract}


\section{Introduction}
\label{s:intro}

The technique of using the white dwarf (WD) sequence in the 
colour-magnitude diagram (CMD) of a star cluster to derive a 
``cooling age'' for the cluster is finally being exploited with 
ever-increasingly success \citep{ric98,hip00,kal01a,and02,han02}. 
This is largely due to the observational power of the Hubble Space Telescope 
(HST) and its ability to detect faint WDs. 
Credit must also go to the development of more sophisticated cooling 
models for WDs (Hansen 1999, for example). 
Strong constraints, independent of cosmological models and parameters, 
are being placed, with this technique, on the age of the Universe. 

The determination of a {\it cooling age} for a star cluster is, as the 
term suggests, based on the fact that WDs cool systematically as they age, 
having only their residual ion thermal energy as a significant 
energy source. 
As a star cluster ages its massive WDs form first from the most massive 
progenitors and 
as time proceeds progressively less massive WDs are introduced at the 
top of the WD cooling track. 
The more massive a WD is the smaller it is (a consequence of being 
supported by electron degeneracy pressure) and the slower it cools. 
The net result is that as a star cluster evolves the WD cooling track 
moves redward in the CMD and that a blue {\it hook} develops as the 
older WDs are caught, in terms of decreasing luminosity, 
by younger (less massive) WDs. 
Furthermore, for any particular WD its cooling rate initially decreases with 
time and this causes the WDs to {\it pile-up} at what is often referred 
to as the {\it bottom} of the cooling track. 
For older populations, such as globular clusters, the oldest WDs will 
actually lie below this point on the CMD because the cooling rate of 
a WD increases sharply in the later stages of its lifetime 
(after $\sim 9 \times 10^9 \,$Gyr depending on mass). 
Regardless we shall continue to refer to this point in the CMD as the 
{\it bottom} of the sequence, which is true at least for open 
clusters\footnote{The description of WD cooling presented here is necessarily 
simplistic. We suggest that the interested reader visit 
{\tt http://astro.ucla.edu/~hansen/m4.html} which includes a clickable WD 
sequence with explanations of the various features.}. 
It is only with HST, and even then with extremely deep exposures, 
that the bottom of the WD cooling sequence in a globular cluster can be 
observed \citep{ric02,han02}. 
For nearby open clusters it is also possible to discover cool WDs by 
conducting deep photometric surveys using ground-based telescopes 
\citep{ric98,kal01a}. 
We note that the {\it blue hook} mentioned above is a mass effect and is 
not that  
described by \citet{han98} in which an individual WD with a hydrogen-rich 
atmosphere evolves blueward in the CMD as it cools below $4000\,$K. 

When calculating the age of a cluster using observations of the WD 
sequence the luminosity function (LF) can be utilised to varying degrees. 
The {\it clump-up} of WDs at the bottom of the cooling track corresponds 
to a maximum, or peak, in the WD LF. 
The presence of this feature in an observed LF for an open cluster is 
evidence that the oldest WDs have been uncovered \citep{hip95}. 
Then, by making assumptions about the mass and composition of these 
WDs, an age is derived by using cooling models to find the time 
taken to cool to the absolute magnitude corresponding to the maximum 
in the LF \citep{ric98,hip00,kal01c}. 
In fact, \citet{bro99} have shown theoretically that a unique relation exists 
between the faintest luminosity of a WD on the cooling sequence of an open 
cluster and the cluster age, as was first suggested by \citet{sch59}. 
For this age to be useful one must be sure that the true maximum in the LF 
has been observed, i.e. the maximum is not produced by incompleteness. 
In the case of a globular cluster, or an open cluster if a large enough 
population of WDs is observed, the more sophisticated method of directly 
fitting the resulting LF can be applied (Andreuzzi et al. 2002). 
As an excellent example of the potential of this technique, \citet{han02} 
have used the WD sequence of the globular cluster M4 to demonstrate a 
clear age difference between this halo object and the Galactic disk. 
A robust lower limit to the age of M4 (and hence the Universe) is found to 
be $12.7 \pm 0.7\,$Gyr. 

An indirect method for using cluster WDs to age the cluster comes 
from matching the WD sequence to the fiducial sequence 
obtained from local WDs with known trigonometric parallax and 
measuring the distance to the star cluster \citep{ren96,zoc01}. 
The distance can then be used to obtain the luminosity of the 
main-sequence (MS) turn-off in the CMD and then an age can be calculated 
by comparison with stellar models. 
The determination of the WD mass fraction in a cluster is also of 
importance as it relates to the nature of the initial mass function (IMF) 
and the WD population of the Galactic halo \citep{hip98}. 

In light of the strong current interest being shown towards WDs in star clusters 
we believe it is pertinent to discuss the theoretically expected behaviour 
of these stars, in particular the morphology of the cluster WD sequence, 
from the point of view of realistic dynamical simulations. 
These same simulations have already been used to demonstrate that significant 
feedback exists between the dense stellar environment of a star cluster 
and the nature of its stellar populations \citep{hur01,hur02,sha02}. 
In this paper we show that the same is true for the cluster WD population 
but that this does not necessarily affect the derived cooling age of a cluster. 
However, we find that extreme care must be taken when using cluster WDs to make 
inferences about the IMF, either when using the WD luminosity function or 
the WD mass fraction. 

In Section~\ref{s:nbsims} we describe the realistic $N$-body models, 
including stellar evolution, used to simulate WD cooling sequences of 
star clusters. 
We discuss the appearance of these sequences in Section~\ref{s:wdseq} and 
investigate the role of mass-segregation in Section~\ref{s:mseg}. 
We then establish to what extent the LF and the mass fraction of the 
WD population is affected by the cluster evolution, in Sections~\ref{s:lumfun} 
and ~\ref{s:wdmfr} respectively, before providing a discussion and 
summary of our findings.

\section{The $N$-body Simulations}
\label{s:nbsims}

\begin{figure}
\plotone{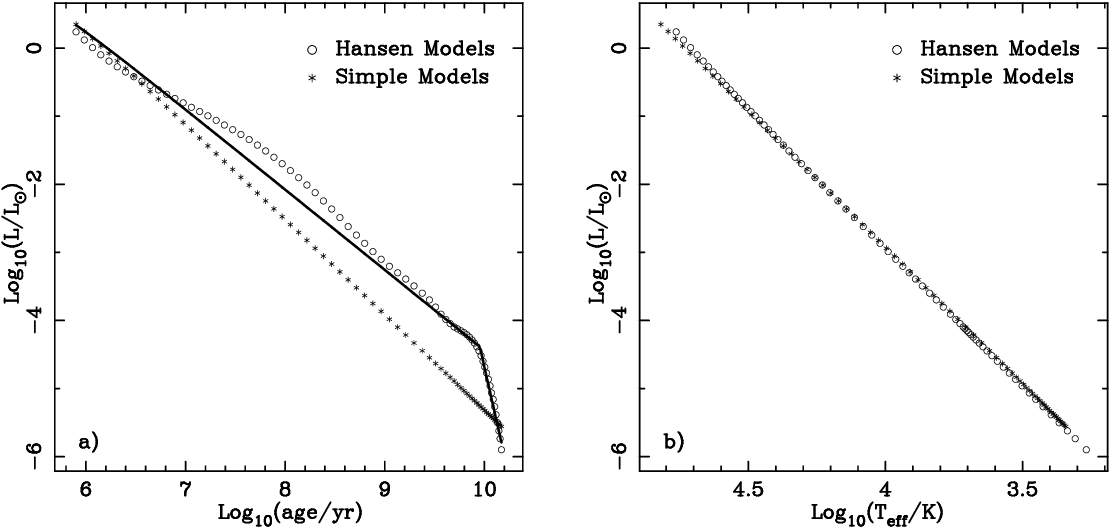}
\caption{
Comparison of the \citet{han99} detailed cooling models with the simple 
cooling track adopted by \citet{hur00}, using a $0.7 M_\odot$ WD as an 
example. 
The left panel shows the evolution of the luminosity with time and 
the right panel shows the Hertzsprung-Russell diagram.  
The radius of the \citet{han99} WD decreases from $\log R/R_\odot = -1.88$ to 
-1.96 in the first $500\,$Myr of cooling and remains approximately constant 
from that point onwards. 
The radius of the \citet{hur00} model is held constant at 
$\log R/R_\odot = -1.94$ throughout.  
Also shown in the left panel (solid line) is the evolution of the 
luminosity with time for the modified-Mestel law used in this paper 
(see text for details). 
Note that at late times the solid line is hidden by the \citet{han99} 
model points. 
\label{f:fig1}}
\end{figure}

We present results from $N$-body simulations performed with the 
Aarseth {\tt NBODY4} code \citep{aar99} on the GRAPE-6 
special-purpose computers \citep{mak98} housed at the American Museum 
of Natural History. 
The {\tt NBODY4} code accounts for the evolution of single stars and 
binaries (mass-loss, mass-transfer, mergers, etc.) 
while modelling all aspects of the dynamical evolution of the cluster 
(see Hurley et al. 2001, and references therein, for full details). 
In particular, the single star evolution algorithm adopted by {\tt NBODY4} 
is that of \citet{hur00}. 
This algorithm models the luminosity evolution of WDs using standard 
cooling theory \citep{mes52} and the radius of a WD is calculated from 
eq.~(17) of \citet{tou97}. 
The stellar evolution algorithm distinguishes three types of WD based 
on the core composition of the giant precursor: 
helium, carbon-oxygen (CO), or oxygen-neon (ONe).  
The CO WDs are 20\% carbon and 80\% oxygen while the ONe WDs are 80\% oxygen 
and 20\% neon (note that neon is representative of all carbon-burning 
products heavier than oxygen). 
Figure~\ref{f:fig1} compares the cooling track of a $0.7 M_\odot$ CO WD 
evolved with the \citet{han99} models with the rather simplified 
model of \citet{hur00}. 
As noted by \citet{ibe84} the Mestel theory assumes that a WD is basically 
an isothermal core composed of an ideal ionic gas embedded in an 
electron-degenerate gas, and that this is surrounded by a thin envelope 
through which photons diffuse at a rate governed by Kramers opacity. 
In Figure~\ref{f:fig1} the decreased cooling rate shown by the \citet{han99} 
WD at early times is explained by the inclusion of neutrino cooling and true 
atmospheric opacities in the detailed models. 
As the WD cools a number of processes neglected by the simple model become 
important: crystallisation of ions in the core, the release of latent heat, 
and the rise of convection as a transport mechanism. 
At late times the crystalline core enters the Debye regime and there is a sharp 
increase in the cooling rate (B. Hansen 2002, private communication). 
To address the differences between the WD cooling rate of the Mestel 
theory and of the \citet{han99} detailed models we have constructed a 
{\it modified}-Mestel cooling law.  
The original Mestel cooling law can be expressed as 
\begin{equation}
L = \frac{b \, M_{\rm WD} Z^{0.4}}
         {\left[ A \left( t + 0.1 \right) \right]^x} \, , 
\end{equation} 
where the factor $b = 635$, the exponent $x = 1.4$, 
$M_{\rm WD}$ is the mass of the WD in solar units, 
$Z$ is the metallicity, $A$ is the baryon number for the WD material, 
and $t$ is the age of the WD in Myr.  
For our modified-Mestel law we split this relation into two 
parts: we use $b = 300$ and $x = 1.18$ for $t < 9\,000\,$Myr, 
and $b = 300 \left( 9\,000 A \right)^{5.3}$ and $x = 6.48$ for 
$t \geq 9\,000\,$Myr. 
As can be seen from Figure~\ref{f:fig1} this rather ad-hoc relation 
provides a much better fit to the detailed models, especially after 
noting that the accuracy of the \citet{han99} models increases for 
$t > 10^8\,$yr (B. Hansen 2002, private communication). 
The modified-Mestel cooling law has been inserted into {\tt NBODY4} 
and is used throughout this work. 
This provides an initial {\it qualitative} assessment of the theoretically 
expected character of cluster WD cooling sequences. 
However, in the near future we will want to directly compare the WD 
sequences and LFs emerging from the $N$-body data with observations, 
and for this it will be important to obtain a more sophisticated fit to 
the behaviour of the realistic WD models 
(as soon as a full database of such models becomes available). 
Metallicity variations are generally neglected when constructing detailed 
WD cooling models because the strong surface gravity of a WD will cause 
all elements heavier than helium to sink below the atmosphere 
\citep{han02}. 
We note that the Mestel cooling law does contain a weak dependence on $Z$ 
through the assumption of electron scattering for the atmospheric opacity. 

We focus on the results of three simulations that each started 
with $28\,000$ stars and a primordial binary frequency, $f_{\rm b}$, of 
40\%, i.e. $12\,000$ single stars and $8\,000$ binaries. 
The IMF of \citet{kro93} was used to assign the masses of single stars 
and a metallicity of $Z = 0.02$ was assumed. 
For primordial binaries the total mass of the binary was chosen from the 
IMF of \citet{kro91}, as this was not corrected for the effect of binaries, 
and the component masses were then assigned according to a uniform 
mass-ratio, $q$, distribution. 
Individual stellar masses were restricted to lie within the limits of 
$0.1 - 50 M_{\odot}$. 
The orbital separation of each primordial binary was taken from the 
log-normal distribution given by \citet{egg89}, within the limits of 
$6 R_{\odot} - 200\,$AU\footnote{The third simulation had an upper limit 
of $50\,$AU but this still exceeds the hard/soft binary limit for simulations 
of this size. According to \citet{heg75} it is only initially hard binaries 
that will contribute to the long-term evolution of the cluster. We note that 
in this case the primordial binary population is effectively representative 
of a larger population drawn from a full range of separations.}, 
and the orbital eccentricity was taken from a thermal distribution 
\citep{heg75}. 
We used a Plummer model \citep{aar74} in virial equilibrium to set the 
initial positions and velocities of the stars but note that the density 
profile quickly evolved to resemble a King model \citep{kin66}. 
The simulated clusters were assumed to be on a circular orbit within a 
Keplerian potential with a speed of $220 \, {\rm km} \, {\rm s}^{-1}$ 
at a distance of $8.5\,$kpc from the Galactic centre.  
Stars were removed from the simulation when their distance from the cluster 
centre exceeded twice the tidal radius defined by this tidal field. 
All stars were on the zero-age main-sequence (ZAMS) when the simulation began 
and any residual gas from the star formation process is assumed to have 
already left the cluster. 
Mass lost from stars during the simulation is simply removed from the cluster, 
with the cluster potential adjusted accordingly. 

Each cluster started with a total mass of $\sim 14\,300 M_{\odot}$ and was 
evolved to an age of $6\,$Gyr at which point $\sim 1\,000 M_{\odot}$ in 
stars remained. 
The initial velocity dispersion of the stars was 
$\sim 3.2 \,{\rm km } \,{\rm s}^{-1}$ (reduced to 
$\sim 1 \,{\rm km } \,{\rm s}^{-1}$ after $6\,$Gyr) and the 
core density was $\sim 500 \,$stars$\,{\rm pc}^{-3}$. 
The average number density of stars in the core throughout the simulations 
was $200\,$stars$\,{\rm pc}^{-3}$. 
The density within the radius that 
contained the inner 10\% of the cluster mass started at the same value, 
reached a minimum of $50\,$stars$\,{\rm pc}^{-3}$ after $\sim 4\,$Gyr, 
and rose to $100\,$stars$\,{\rm pc}^{-3}$ at $6\,$Gyr. 
Each simulated cluster showed a modest core-collapse at $\sim 1\,$Gyr, 
identified by a significant increase in core density, but we note that 
for models with such a large proportion of primordial binaries the exact 
point of core-collapse is difficult to judge, if it occurs at all. 

For reference purposes we have performed two additional simulations 
that each started with $28\,000$ stars but no primordial binaries. 
We also draw upon the simulations with $20\,000$ stars and $f_{\rm b} = 10\%$ 
described by \citet{sha02} in their work on double-WD binaries in star 
clusters. 
As a result, our findings are directly applicable to intermediate-mass and 
massive open clusters. 
By making the appropriate scalings we can 
(cautiously) make inferences relating to WDs in globular clusters as well. 
More realistic, $N \geq 10^5$, simulations will be needed to make definite 
predictions about globular clusters.

\section{The White Dwarf Sequence}
\label{s:wdseq}

We have chosen to concentrate on the simulated cluster data 
at $4\,$Gyr for the purpose of illustrating the nature and appearance 
of the cluster WD population. 
There are a number of reasons for this choice. 
Firstly, this is late enough in the simulation (approximately 12 half-mass 
relaxation times have elapsed) that the cluster is dynamically evolved. 
On the other hand, it is not so late that the number counts of the stellar 
populations have become statistically insignificant. 
Furthermore, if time is scaled by the half-mass relaxation timescale, 
$t_{\rm rh}$, then 
an age of $4\,$Gyr for a cluster of $\sim 30\,000$ stars is representative 
of a $100\,000$ star cluster at $12\,$Gyr \citep{mey97}, 
{\it i.e. the results of our massive open cluster simulations at $4\,$Gyr 
can be related to a moderate size globular cluster}. 
We regard this final point with some caution because a number 
of timescales are at work in a star cluster and these scale differently 
with $N$ \citep{aar98}. 
It has also been shown that many of the structural properties of a star 
cluster are $N$-dependent \citep{goo87}. 

\begin{figure}
\plotone{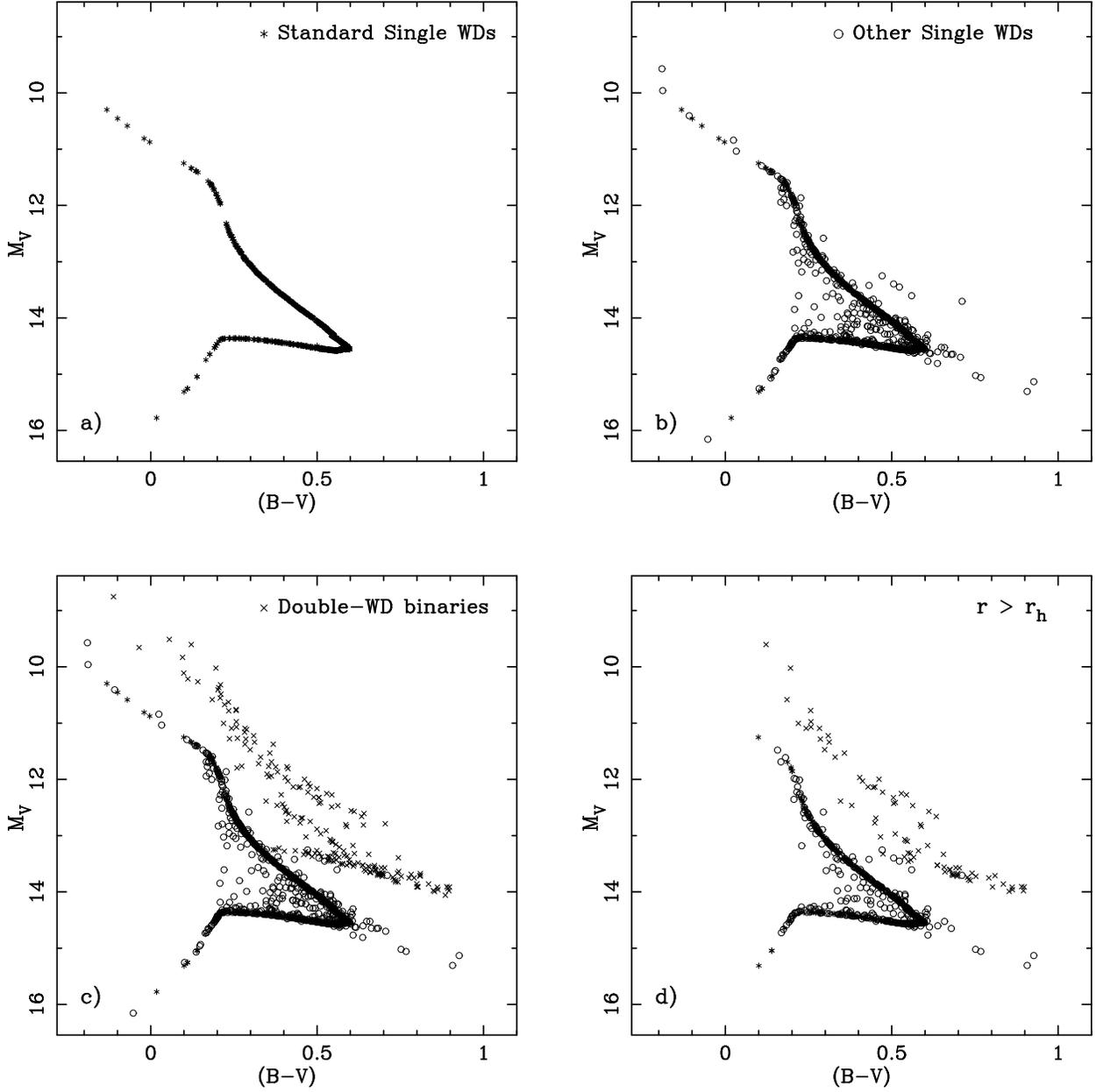}
\caption{
Cluster CMD for WDs at $4\,$Gyr. 
Stars from three $N$-body simulations, each with $N_0 \simeq 28\,000$ 
and $f_{\rm b} \sim 40\%$, are shown.  
All WDs are assumed to be of DA type and bolometric corrections are 
taken from \citet{ber95}. 
We have distinguished three different types of WDs depending on their 
binarity and formation path: single WDs that evolved from single stars 
(standard), single WDs for which the progenitor star (or stars) 
was previously the member of a binary, and double-WD binaries. 
Note that all binaries are assumed to be unresolved. 
Panel~(a) shows only the standard WDs, panel~(b) adds in the remaining 
single WDs, and panel~(c) shows all three types. 
Panel~(d) is a replica of~(c) but shows only WDs that lie outside of 
the cluster half-mass radius (typically $4.5\,$pc).  
There are a total of 863 standard WDs, 598 single WDs that evolved via 
a binary phase, and 198 double-WD binaries (25\% of these formed via an 
exchange interaction). 
\label{f:fig2}}
\end{figure}

Figure~\ref{f:fig2} shows the CMD at $4\,$Gyr for all WDs in the three 
$N = 28\,000$ simulations with primordial binaries. 
To convert the theoretically derived quantities of luminosity and 
effective temperature to magnitudes and colours we have used the 
bolometric corrections provided by the WD models of \citet{ber95}. 
We start in Figure~\ref{f:fig2}a by plotting only what we call standard 
single-WDs. 
By this we mean that each of the WDs, and their progenitor stars, 
were never part of a binary or involved in a collision and have evolved according 
to the standard picture of single star evolution. 
This produces the smooth cooling track seen in Figure~\ref{f:fig2}a.  
For these WDs it is true that the more advanced along the track that a 
particular WD is, the more massive and older it is. 
We highlight the standard single WDs because these are the objects that the 
cooling models used to age the WD sequence directly relate to. 

Next we add in all the remaining WDs that are single at $4\,$Gyr but whose 
progenitor was originally a member of a binary 
(Figure~\ref{f:fig2}b). 
Even though a substantial fraction of these WDs overlie the standard 
cooling track it is evident that the remainder contribute a great deal of 
scatter to the CMD. 
As an example, some of these WDs have evolved from blue stragglers, 
or more generally any MS star rejuvenated by mass-transfer. 
These progenitors' journeys to the asymptotic giant branch have been delayed, 
hence when the WD was born it was more massive than WDs born from standard 
single stars at that time. 
This yields WDs lying below the standard WD sequence. 
Conversely, WDs less massive than expected at birth are produced 
from giants initially in binaries that overfilled their Roche-lobe and lost 
their envelopes prematurely, and then lost their partners in exchange 
interactions. 

\begin{figure}
\plotone{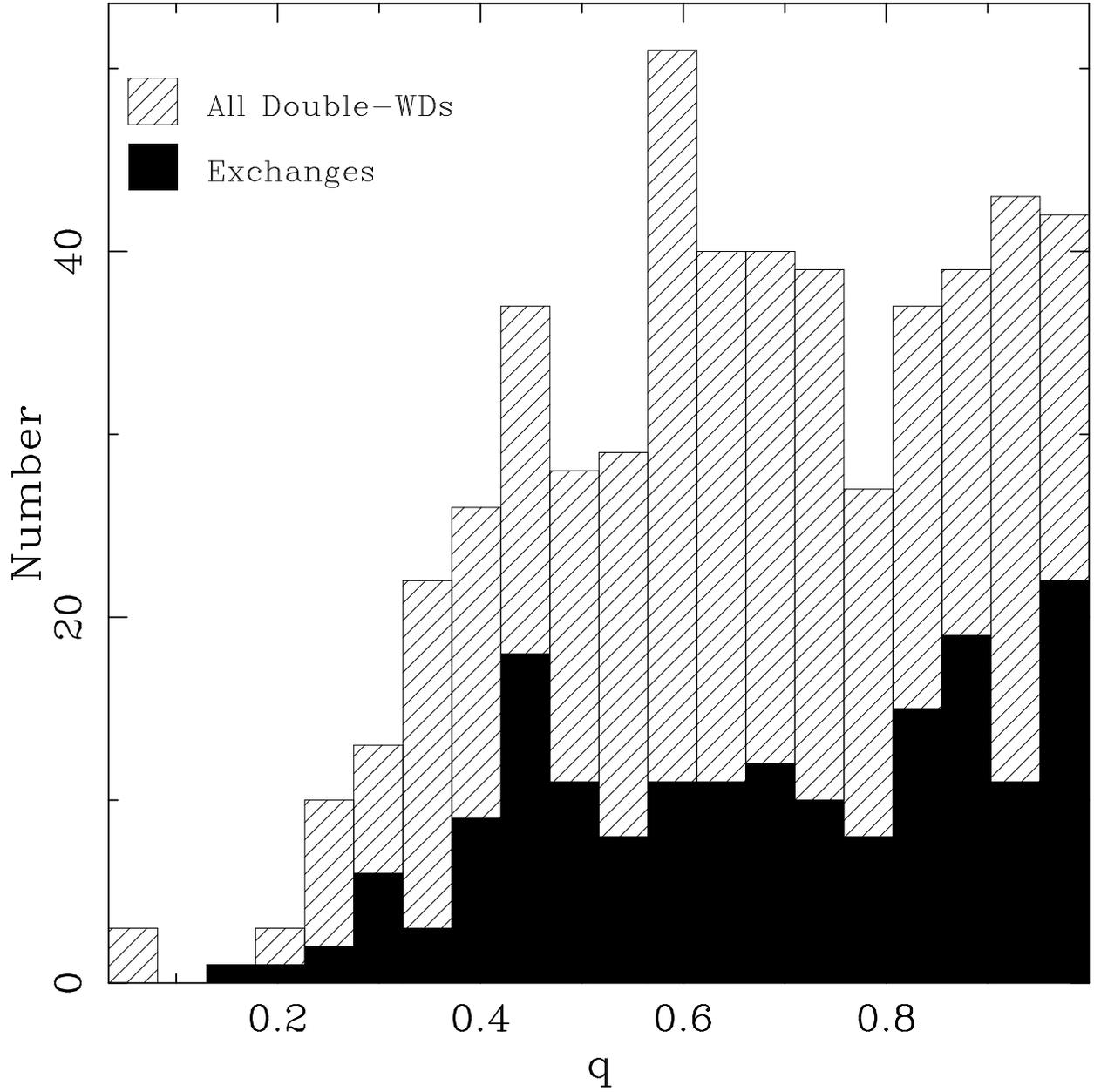}
\caption{
Mass-ratio, $q$, distribution of double-white-dwarfs present in an 
$N$-body simulation at $4.0\,$Gyr. 
Note that we define $q$ so that it is always less than unity. 
\label{f:fig3}}
\end{figure}

In Figure~\ref{f:fig2}c we complete the full WD CMD by including all 
the double-WD binaries present at $4\,$Gyr. 
The first thing to notice is that for the most part the double-WD 
sequence is well separated from the standard cooling track. 
This is because the computed double-WD binaries are mainly high mass-ratio 
systems (see Figure~\ref{f:fig3}), in agreement with the measured 
mass-ratios of local double-WDs \citep{max02}. 
Provided that photometric errors are modest (say, $\leq 0.5\,$mag) 
it is possible to minimize contamination of the WD sequence by double-WDs 
in at least the upper half of the WD CMD. 
Further down the sequence we notice that double-WDs clump-up at brighter 
magnitudes than do the single-WDs and start to approach the 
WD sequence. 
This, combined with the scatter produced by the non-standard single WDs, 
leads to our first note of caution regarding observations of cluster 
WD sequences. 
{\it Estimating the position of the bottom of the cooling track by the detection of 
WDs blueward of the track, or by a build-up of WDs at a certain magnitude, 
can be seriously misleading.} 
It is possible to just be seeing the scatter in the WD sequence produced by 
non-standard WD evolution, or by a population of old double-WDs, and one may 
need to go deeper to find the true extent of the track. 
If the termination of the double-WD sequence were to be mistakenly used as 
the bottom luminosity of the WD sequence then, using the relation given by 
\citet{bro99}, this would translate to underestimating the cluster age by 
$1.6\,$Gyr for a cluster with an actual age of $4\,$Gyr. 

In producing the WD CMD we have assumed that all the WDs are hydrogen-line 
(DA) type whereas a small, but significant, fraction will actually have 
strong helium (DB), or other anomalous, features.  
At least 75\% of spectroscopically identified WDs are classified as 
DA in the catalogue of McCook \& Sion (1999) but we note that for cool 
WDs the DA:DB ratio is more likely 1:1 \citep{ber97,han99}.   
Modelling of WDs that do not have pure hydrogen atmospheres will produce 
additional scatter in the WD CMD, as will other factors that 
affect the cooling times and temperatures of WDs:  
the relative fractions of carbon and oxygen in the interiors of CO WDs 
\citep{koe02}, and the mass of the hydrogen (or helium) envelope 
\citep{han99}, for example. 
Importantly, our simulations have been performed with moderate 
stellar density - at least an order of magnitude less than conditions 
within the core of an actual globular cluster - and at higher density the 
incidence of stellar interaction is expected to be higher and hence a greater 
number of non-standard single WDs will be produced\footnote{For example, the 
high density globular cluster M80 is observed to have a large number of blue 
stragglers in its core (305; Ferraro et al. 1999)  
and therefore would be expected to also contain a large 
number of non-standard WDs.}. 
Counteracting this, globular clusters are observed to have smaller 
binary fractions than used in our simulations: M4 has a binary frequency 
that could be as high as 15\% \citep{cot96} or as low as 4\% \citep{ric96}. 
It is also possible that a higher stellar density may hinder the production 
of certain populations, such as double-WDs, and lead to less contamination 
of the WD sequence. 

Outside of the half-mass radius, $r_{\rm h}$, of a star cluster the 
number density of stars is less than in the core and the incidence 
of stellar interactions is also less. 
The binary fraction is also smaller in this region as mass-segregation 
is effective in causing binaries to sink towards the cluster centre 
(see Section~\ref{s:mseg}). 
These considerations lead to a much cleaner WD sequence, as shown in 
Figure~\ref{f:fig2}d, and it is here that observations of WD sequences 
(for the purposes of age dating) in globular clusters can most cleanly 
be conducted, if enough WDs are present. 

\begin{figure}
\plotone{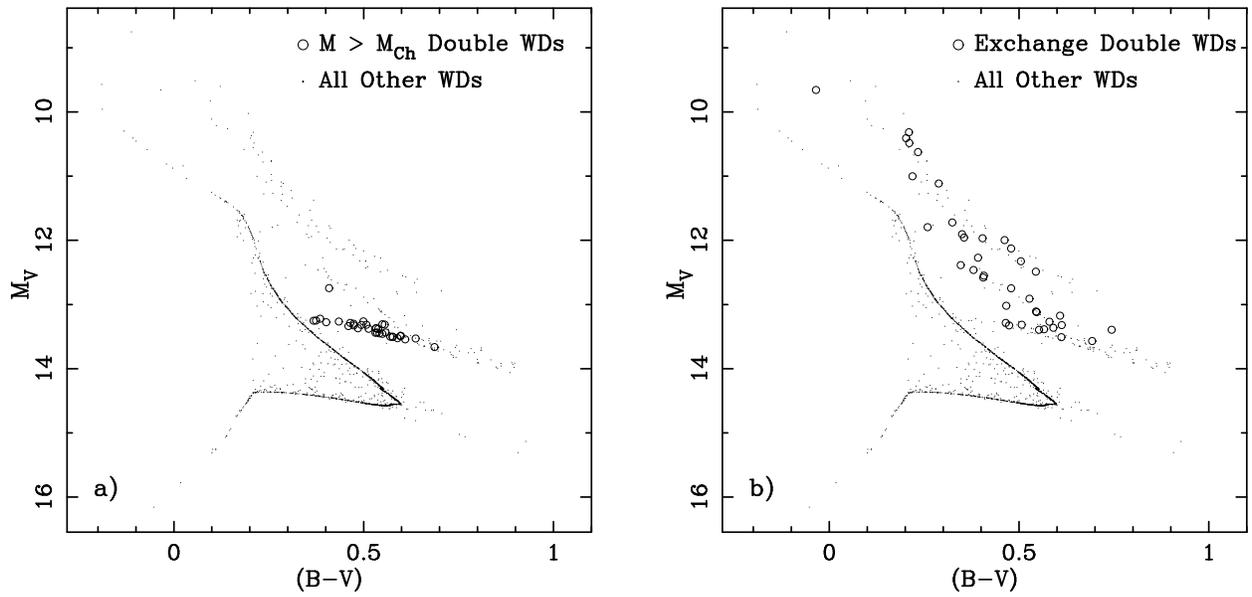}
\caption{
Same as Figure~\ref{f:fig2}c but in this case we distinguish 
double-WD binaries that have a combined mass in excess of the 
Chandrasekhar limit for a single WD (left panel), and double-WD 
binaries that formed via an exchange interaction (right panel). 
\label{f:fig4}}
\end{figure}

\citet{sha02} found that open star clusters produce supra-Chandrasekhar mass 
double-WD binaries with merger timescales less than a Hubble time at a 
greatly enhanced rate relative to the field. 
Of the 198 double-WDs shown in Figure~\ref{f:fig2}, 35 have a combined mass 
in excess of the Chandrasekhar mass and these are highlighted in 
Figure~\ref{f:fig4} (left panel). 
All but 2 of the supra-Chandrasekhar mass double-WDs lie in a clump just 
to the right of the standard single-WD sequence. 
The area defined by this clump also contains 12 other stars (non-standard 
single WDs and sub-Chandrasekhar mass double-WDs) which means that 73\% 
of the stars in this sub-area of the CMD are supra-Chandrasekhar mass 
double-WDs. 
Provided that observations can be performed with suitably high signal-to-noise 
this method is a possible way to isolate these potentially interesting 
binaries. 
More sophisticated follow-up methods, such as the use 
of a gravitational wave detector \citep{ben99}, would be required to 
learn about the merger timescales of these binaries. 
Figure~\ref{f:fig4} (right panel) also highlights the double-WD binaries 
produced by exchange interactions during the simulations: 25\% of the 
double-WD binaries at $4\,$Gyr are non-primordial. 
It can easily be seen that these binaries do not preferentially form in any 
particular sub-area of the general double-WD sequence 
(see also Figure~\ref{f:fig3}) and hence they cannot be isolated by photometric 
methods.

\section{Mass Segregation}
\label{s:mseg}

\begin{figure}
\plotone{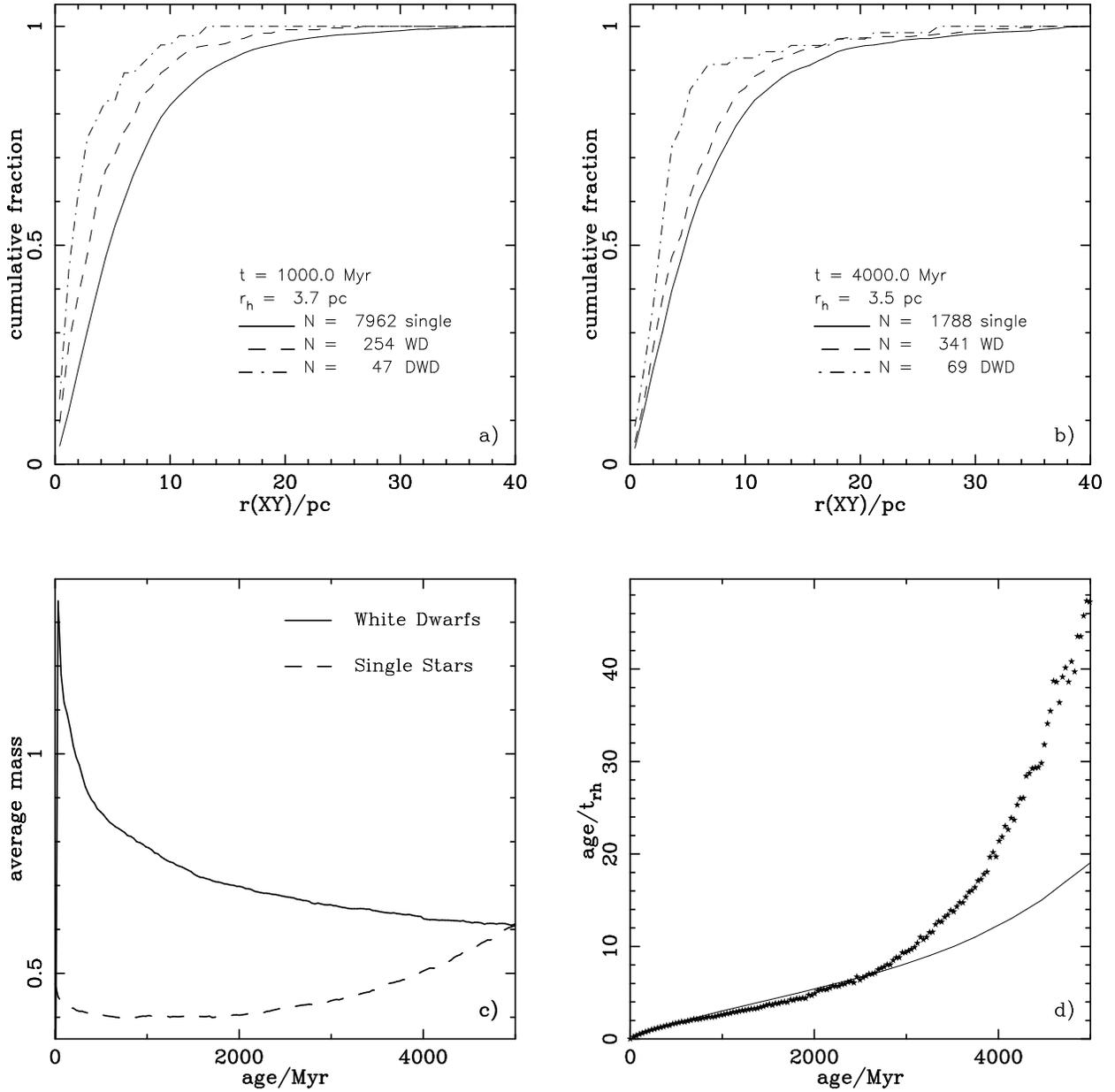}
\caption{
The top panels show the cumulative radial distributions of single stars, 
single WDs, and double-WD binaries at ages of $1\,$Gyr (top left) and 
$4\,$Gyr (top right). 
Also shown, as a function of time, is the evolution of the average 
stellar mass for WDs and non-WD single stars (bottom left), 
and the cluster age scaled by the half-mass relaxation timescale, 
$t_{\rm rh}$, current at that time (bottom right, 
the solid line represents the number of actual half-mass relaxation 
times elapsed by using the integrated half-mass relaxation timescale). 
Data from the three simulations with $N \sim 28\,000$ and 
$f_{\rm b} \sim 40\%$ are included. 
\label{f:fig5}}
\end{figure}

A number of $N$-body studies \citep{gie97,por01,hur01,hsh02} have 
previously verified the Fokker-Planck results of \citet{che90}:  
mass-segregation occurs in star clusters, and this causes stars 
less massive than the average stellar mass to migrate outwards on a 
timescale governed by two-body relaxation. 
Conversely, stars more massive than average sink towards the centre of 
the cluster. 
Low-mass stars are thus preferentially stripped from the cluster 
by the external potential of the Galaxy. 
To reinforce these findings we show in Figures~\ref{f:fig5}a and ~\ref{f:fig5}b 
population gradients for single stars, WDs, and double-WDs, at $1$ and $4\,$Gyr 
respectively. 
Clearly the WDs are more centrally concentrated than the overall population 
of single stars and therefore it is less likely that WDs will be lost from the 
cluster by tidal stripping. 
In fact, for our $N = 28\,000$, $f_{\rm b} = 40\%$ simulations after 
$4\,$Gyr of evolution, 20\% 
of the mass generated in WDs has escaped from the cluster with the average 
mass of these WD escapers being $\sim 0.7 M_\odot$. 
By contrast, 91\% of the mass in single MS stars with $M < 0.7 M_\odot$ has 
been lost from our models after the same period of time. 
After $1\,$Gyr only 4\% of the WD mass has escaped from the cluster. 
The fraction of mass lost from the cluster in escaping WDs agrees 
favourably with \citet{ves97} when comparing with their model 
at $8\,$kpc from the Galactic centre after a similiar number of relaxation 
times have elapsed. 

Some central concentration of the WD population is to be expected because 
their average mass is greater than that of all the cluster stars 
(for $t < 5\,$Gyr; see Figure~\ref{f:fig5}c). 
However, the main reason for this concentration is that the progenitors 
of the WDs were originally 
more massive than the current MS turn-off mass and therefore the WDs are 
more likely to be born interior to $r_{\rm h}$.  
This point was also discussed by \citet{por01} in relation to their 
$N = 3\,000$ models of young open clusters. 
Binaries are on average more massive than single stars and as such will 
segregate towards the centre of the cluster.  
This is also true of double-WD binaries (see top panels of Figure~\ref{f:fig5}) 
which have an even higher average mass than standard binaries. 

Figure~\ref{f:fig5}c shows the average mass of the WDs as a function of time, 
and also the average mass of all the single stars in the cluster, excluding 
WDs and other degenerate objects. 
The single star average mass initially decreases owing to mass-loss from the 
most massive stars but then begins to increase as tidal stripping of low-mass 
stars slowly starts to dominate over mass-loss from stellar evolution. 
If the cluster was instead evolved in isolation so that tidal stripping was 
not accounted for, then the single star average mass at $4\,$Gyr would be 
$0.36 M_\odot$ as opposed to $0.51 M_\odot$. 
The reverse is true for the evolution of the WD average mass -- stellar evolution 
and not tidal stripping is the dominant factor. 
The first WD forms at $\sim 40\,$Myr from a star with a ZAMS mass of $8 M_\odot$ 
and is of oxygen-neon composition with a mass slightly below the 
Chandrasekhar mass ($1.44 M_\odot$). 
Then, as time proceeds, the zero-age WD mass progressively decreases 
(see Hurley, Pols \& Tout 2000 for a full description of the WD initial-final 
mass relation generated by the evolution algorithm) which eventually results 
in a build-up of CO WDs with masses in the range $0.6 - 0.7 M_\odot$. 
Helium composition WDs are also produced, but only after some form of 
binary interaction. 
These have an average mass of $0.34 M_\odot$.  
The average WD mass at $4\,$Gyr is $0.62 M_\odot$, practically the same as it 
would be if the population had been evolved in isolation. 
While the single star average mass is sensitive to where in the cluster it is 
measured, decreasing from $0.82 M_\odot$ in the core to $0.37 M_\odot$ near 
the tidal boundary at $4\,$Gyr, the WD average mass is relatively uniform 
throughout the cluster. 
Thus it would appear that overall the WD population has been little affected 
by the dynamical evolution of the cluster, in agreement with the conclusion 
of \citet{por01}, although we will show in the next section that this is not 
necessarily true when considering the WD LF. 

The fact that over time the progenitor mass of the WDs is steadily decreasing, 
and that the WD and single star average masses are nearing equality, 
explains why the WDs appear marginally less centrally 
concentrated at $4\,$Gyr than at $1\,$Gyr. 
Eventually the average mass of the non-degenerate single stars will increase 
above that of the WDs and the population gradients of the two populations will 
converge -- mainly due to the WDs drifting slowly outwards. 
However, by the time this occurs the cluster is dynamically very old, in excess 
of 20 half-mass relaxation times will have elapsed, and near complete disruption. 

In Figure~\ref{f:fig5}d we show how the cluster age scales with the 
half-mass relaxation timescale as this can be a useful reference tool when 
interpreting the results of the simulations. 
We show both the number of half-mass relaxation times elapsed when simply 
dividing the cluster age by the current $t_{\rm rh}$ and when integrating 
$t_{\rm rh}$ over the lifetime of the cluster. 
The latter number is representative of the true dynamical age of the cluster 
and the two methods give similar results until $t_{\rm rh}$ starts to 
decrease during the latter stages of evolution. 
The half-mass relaxation timescale for the simulations starts at 
$\sim 200\,$Myr, rises to $\sim 450\,$Myr at $2\,$Gyr, and has decreased 
back to $\sim 200\,$Myr after $4\,$Gyr of evolution. 
It basically follows the evolution of the cluster half-mass radius 
which increases initially owing to mass-loss from massive stars 
and then decreases as it starts to feel the effect of the shrinking 
tidal radius \citep{hur01}. 
At $4\,$Gyr, $r_{\rm h} = 4\,$pc for these simulations.

\section{The Luminosity Function}
\label{s:lumfun}

The WD luminosity function holds information about the age of a 
cluster and the IMF of its stars. 
In order to extract this information accurately we must be sure that 
the observed WD LF is relevant for the intended purpose. 
For the age determination this really boils down to being certain that 
the true peak in the LF has been identified. 
In the case of open clusters this corresponds to being certain that the 
{\it bottom} of the WD sequence has been reached, 
i.e. the coolest standard single WD has been observed. 
However, for the older populations of globular clusters it is not yet possible 
to observe the oldest WDs as these will be massive helium atmosphere WDs with 
luminosities below current detection limits \citep{han99}. 
The attraction of this approach is that it provides an age that is 
relatively independent of stellar evolutionary 
models because the nuclear-burning lifetime of the progenitor to the 
oldest WDs is short ($\sim 40\,$Myr) compared to the age of all globular 
clusters and most open clusters \citep{hip95}. 
For the case of inferring the IMF we must be sure that the observed 
WD LF is a true representation of the present-day mass function (PDMF) 
of the cluster, 
noting that for WDs the LF and MF are directly related via the cooling 
models. 

\begin{figure}
\plotone{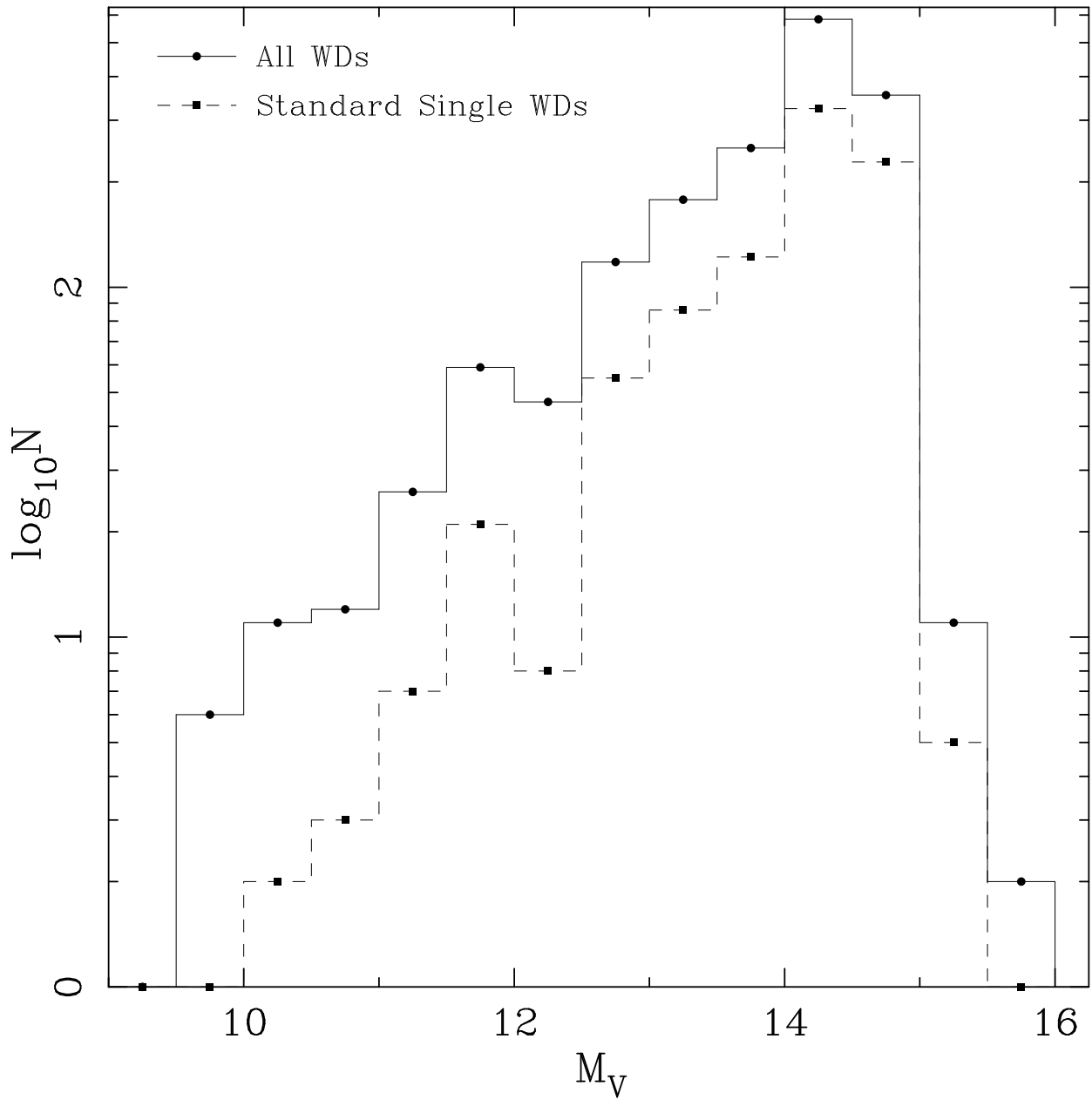}
\caption{
WD luminosity functions at $4\,$Gyr for all WDs (solid line), and  
for single WDs that evolved from single stars (dashed line). 
WDs from all three simulations with $N \sim 28\,000$ and 
$f_{\rm b} \sim 40\%$ are included. 
The LF for all WDs (which includes double-WD binaries) 
is best fit across the range of masses that produce 
WDs for $t \leq 4\,$Gyr ($1.4 - 8.0 M_{\odot}$ for $Z = 0.02$) 
by a \citet{sal55} IMF with slope $\alpha = 3.75$. 
The LF for standard single WDs is best fit by a \citet{sal55} IMF with slope 
$\alpha = 3.15$. 
\label{f:fig6}}
\end{figure}

In Figure~\ref{f:fig6} we show the WD LF from the $N$-body simulations 
at $4\,$Gyr for the entire WD sequence (as shown in Figure~\ref{f:fig2}c). 
We also show the LF for the standard single-WDs.  
Performing a $\chi^2$ test reveals a probability of 0.02 that the two LFs 
are drawn from the same distribution. 
The main points to notice are that contamination of the WD sequence by the 
presence of double-WDs and non-standard single-WDs does not affect the 
location of the LF maximum but that the slope of the LF {\it is strongly} 
affected. 

\subsection{Inferred IMF Slopes}
\label{s:imfslp}

For a population with $Z = 0.02$ at an age of $4\,$Gyr the range of 
ZAMS stellar masses that will have evolved to become 
WDs is $1.4 - 8.0 M_\odot$. 
Assuming a power-law IMF of the \citet{sal55} form (where $\alpha = 2.35$ 
is the corresponding slope of the function in our chosen notation) we 
can fit the IMF slope in this mass range to the cluster LF by constructing 
theoretical LFs for a range of $\alpha$. 
In this case the LFs from the $N$-body simulations are the {\it observed} 
LFs. 
When constructing the theoretical LFs we evolve only single stars as we 
want to quantify how the inclusion of non-standard WDs in the LF affects 
the inferred IMF. 
These single stars are evolved according to the same rapid evolution algorithm 
used by {\tt NBODY4}. 
This ensures that uncertainties in the accuracy of the WD cooling tracks 
and the MS lifetimes of the WD progenitors, for example, 
do not play a role in the fitting process. 
No dynamical effects are accounted for in the theoretical LFs. 
To determine the appropriate $\alpha$ for an observed LF we find the 
theoretical LF that gives the smallest value of the $\chi^2$ statistic,  
where we have used $\Delta \alpha = 0.05$ in constructing the theoretical LFs. 
The two distributions are normalized so that the sum of all bins are 
equal which means that the number of degrees of freedom in the fitting 
process is one less than the number of bins. 
We note that the probability returned by the $\chi^2$ fit is not independent 
of the normalization -- if the distributions are normalized to a greater 
total number of WDs the probability that the fit is a good one decreases 
(for the same number of bins). 
In this work we are primarily concerned with using the value of $\chi^2$ 
to determine which $\alpha$ gives the best fit. 
However, the probability returned by the fit may also be of interest so 
we have made sure to use the same normalization throughout this work, 
i.e. all distributions are normalized to have the same total number of WDs 
as that given by the solid line in Figure~\ref{f:fig6} before commencing 
the fitting process. 

The result is $\alpha = 3.75$ (with a probability of 0.30 that this is 
a good fit) when considering all the WDs, and $\alpha = 3.15$ (0.65) 
for the standard single-WD LF.  
We note that the slope of the \citet{kro93} IMF used in these simulations 
is 2.7 for masses in excess of a solar mass and therefore, after $4\,$Gyr 
of evolution, the standard single-WDs are no longer representative of the 
true PDMF for the initial population (evolved without dynamics). 
After $1\,$Gyr the LF for standard single-WDs is best fit by 
$\alpha = 2.70$ (0.87) and after $2\,$Gyr the fit reveals 
$\alpha = 2.80$ (0.76) -- further evidence that over time the cluster 
environment is eroding the usefulness of the WDs as tracers of the IMF 
but also showing that it takes time for this erosion to become 
significant. 
The presence of double-WDs in the LF increases the relative number counts 
in the intermediate-magnitude bins (located in the range of $\sim 1 - 3$ 
mags below the peak), effectively {\it puffing-up} the middle of the LF 
and causing the fitting process to find an artificially high $\alpha$ 
as the IMF tries to produce more single-WDs of intermediate age. 
The error in the fitting process is understandably higher in this case as 
it becomes difficult to fit the LF with a single power-law IMF. 
When including the 40\% binary fraction in the initial population its 
non-dynamical PDMF at $4\,$Gyr is best fit with $\alpha = 3.15$ (0.48). 
After $1\,$Gyr the LF for all cluster WDs, including double-WDs, is best fit by 
$\alpha = 2.75$ (0.93) and after $2\,$Gyr it is best fit by 
$\alpha = 3.00$ (0.43). 
So the LF for all WDs is affected by the cluster environment at a much earlier 
stage than for the standard single-WDs, first indicating a flatter than expected 
mass function which then becomes steeper than expected at later times. 

\subsection{Half-mass Radius}
\label{s:rhalf}

\begin{figure}
\plotone{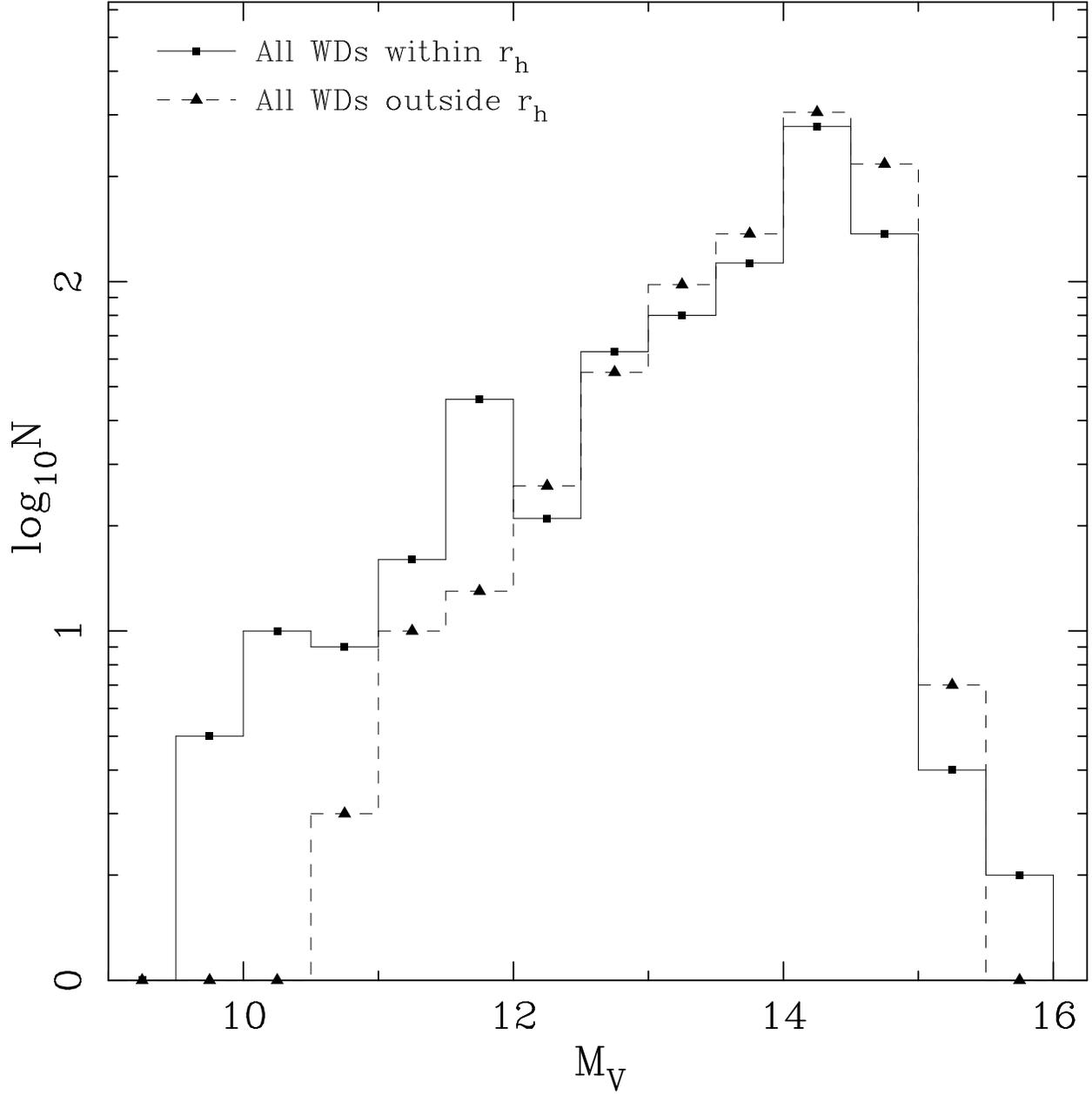}
\caption{
WD luminosity functions at $4\,$Gyr for all WDs interior (solid line) 
and exterior (dashed line) to the cluster half-mass radius. 
The best fitting \citet{sal55} IMFs for each case have slopes of 
$\alpha = 3.90$ and $\alpha = 3.55$, respectively. 
\label{f:fig7}}
\end{figure}

In Figure~\ref{f:fig7} we break the LF for all WDs 
at $4\,$Gyr into two separate LFs depending on whether the WD is 
inside or outside of the cluster half-mass radius. 
Performing a $\chi^2$ test on the two LFs reveals a probability of 0.01 that 
they are drawn from the same distribution. 
The inferred IMF slope for $r < r_{\rm h}$ is $\alpha = 3.90$ (0.86) and 
exterior to $r_{\rm h}$ it is $\alpha = 3.55$ (0.26), which makes sense on the 
basis of mass-segregation increasing the proportion of double-WDs and 
luminous single-WDs in the central regions (see also Figure~\ref{f:fig2}d). 
Considering only the standard single-WDs the best fits are $\alpha = 3.30$ (0.94) 
for $r < r_{\rm h}$ and $\alpha = 3.00$ (0.46) for $r > r_{\rm h}$.  
The results of the LF fits at 1, 2, and $4\,$Gyr for the simulations 
starting with $N = 28\,000$ and $f_{\rm b} = 40\%$ are summarised in 
Table~\ref{t:table1}. 

It is clear from Figure~\ref{f:fig7} that the most luminous WDs are 
preferentially found interior to $r_{\rm h}$. 
We find that when a WD appears at the top of the cooling track it is 
most likely to be found at a radial distance of $0.6 r_{\rm h,WD}$ 
from the cluster centre, 
where $r_{\rm h,WD}$ is the half-mass radius of the WDs which is 
itself less than $r_{\rm h}$ (for $t < 5\,$Gyr). 
Then as the WD cools it will relax out to $r_{\rm h,WD}$ on a timescale 
equivalent to twice the current half-mass relaxation timescale. 
By the time the WD reaches the bottom of the cooling track it will have 
relaxed even further to blend in with the spatial distribution of the 
cool WDs. 
As the cluster evolves the fraction of luminous WDs found interior to 
$r_{\rm h}$ decreases. 
Defining a luminous WD as one appearing in the upper four magnitudes of 
the WD sequence ($M_V < 14$ at $4\,$Gyr) we find that 19\% of the 
standard WDs with $r < r_{\rm h}$ at $4\,$Gyr are luminous whereas 
40\% are luminous at $1\,$Gyr. 
Exterior to $r_{\rm h}$ the numbers are 14\% and 24\% which means that 
over time the difference between the two populations is also decreasing. 
This is apparent in Table~\ref{t:table1} where we see that the difference 
between the LFs for WDs interior and exterior to $r_{\rm h}$, in terms of 
the $\alpha$ fitted to each LF, decreases as the cluster ages.  

\subsection{Exchanges and Escapers}
\label{s:excesc}

The reason for the steepening with time of the IMF inferred from the standard 
single-WD population of the cluster is not an over-abundance of luminous 
WDs but rather the preferential escape of older less-luminous WDs from the 
cluster, as these are more likely to reside in the outer regions. 
We recall from the previous section that the fraction of escaping WDs 
increases as the cluster evolves. 
Strong dynamical interactions with other cluster stars also alters the 
make-up of the standard single-WD population but here the effect on the LF 
is less clear. 
After $4\,$Gyr 14\% of the potential standard single-WDs have been lost 
from the population because they, or their MS star or giant precursors, 
have been exchanged into binary systems\footnote{ 
The approximate timescale \citep{dav95} for a $1.4 M_\odot$ MS star to be 
exchanged into a binary consisting of $1.0$ and $0.5 M_\odot$ MS stars is 
$1\,$Gyr if the binary is in the core of the cluster and has a semi-major axis 
of $10\,$AU (or equivalently if the binary is at $r_{\rm h}$  
and has a semi-major axis of $50\,$AU).}.  
This accounts for 49\% of the total number of exchange interactions 
recorded in the $4\,$Gyr of cluster evolution. 
In raw numbers this amounts to a loss of 140 potential standard single-WDs: 
19 of these were exchanged into a binary after becoming a WD and 12 of the 
double-WDs present at $4\,$Gyr contain a WD that would have been a 
standard single-WD without dynamical intervention. 
At any point in time this process is more likely to affect luminous WDs as 
these reside in higher density regions but over the lifetime of the 
cluster both cool and hot WDs are affected and the change to the LF slope 
is minimal. 
After $1\,$Gyr of cluster evolution 6\% of the standard single-WD population 
has been lost owing to exchange interactions (8\% after $2\,$Gyr), 
accounting for 24\% of all exchanges to that point (37\% after $2\,$Gyr), 
and none are found in double-WDs. 
Therefore, as the cluster evolves the standard single-WDs become involved 
in dynamical encounters to a greater degree but the effect remains 
secondary to the escape of cool WDs in explaining the steepening of the 
inferred IMF. 

When considering all WDs appearing in the cluster WD sequence, 
including double-WDs, the LF at $1\,$Gyr has suffered from a decrease in 
the relative number counts at the bright end (as evidenced by the lower 
than expected $\alpha$). 
There are a number of ways that the cluster environment could be producing 
this effect. 
The presence of wide double-WD binaries in the core leaves this population 
exposed to disruption by 3- and 4-body encounters. 
The presence in the core of binaries composed of a WD and a nuclear burning star 
that will soon evolve to become a WD, raises the possibility of 3-body 
interactions exchanging the WD for a slightly more massive MS star so that 
this system will not reach the WD sequence. 
Furthermore, the progenitors of short-period double-WDs, that are formed via a 
common-envelope phase, may be hardened by 3-body encounters and as a result 
the common-envelope is formed earlier than expected, thus accelerating the 
formation of the double-WD, or even resulting in a merger event.  
The time at which the value of the IMF slope inferred from the LF of all 
WDs matches the expected non-dynamical value ($\alpha = 3.15$) is 
$\sim 2.4\,$Gyr. 
By this time the density of stars in the inner regions of the cluster was 
approximately a factor of 10 less than the starting value, making dynamical 
modification of the cluster populations less likely, and the escape of cool 
WDs had started to dominate in terms of shaping the LF. 

\subsection{Where to Observe?}
\label{s:obclue}

\citet{bro99} have already demonstrated that variations to the IMF do not 
alter the location of the LF maximum and therefore calculations of the 
cluster age using this method are not sensitive to errors in the IMF 
slope. 
However, for the more sophisticated method of fitting the entire LF 
(Hansen et al. 2002, for example) the method {\it is} sensitive to 
the IMF as it alters the relative number of WDs in each luminosity 
(or mass) bin. 
Working on the premise that we are interested in deriving an age for a 
star cluster by fitting the WD LF, and that we are only going to consider 
cooling models of standard single-WDs in this process, then the question 
is this: where in the cluster should we look in order to extract the 
correct PDMF, i.e. an IMF slope of $\alpha = 2.7$ for the single-WDs? 
The simple answer for our models at $4\,$Gyr is: {\it nowhere}. 
At this point the cluster is dynamically well evolved (12 half-mass 
relaxation timescales have elapsed) and the cluster environment has 
effectively removed all trace of the IMF from the WD population. 
The region where we come closest to recovering the PDMF is 
$1 < r/r_{\rm h} < 2$ where $\alpha = 2.95$ for the standard single-WDs. 
At earlier times the situation is not so severe. 
After $1\,$Gyr (3 half-mass relaxation timescales) and $2\,$Gyr 
(5 half-mass relaxation timescales) the PDMF for single-WDs is 
recovered by considering only 
standard single-WDs in the region $0.5 < r/r_{\rm h} < 1.0$. 
Owing to the fact that the WDs are centrally concentrated with respect 
to nuclear-burning single stars at these times 
(see top panels of Figure~\ref{f:fig5}) 
it is understandable that we need to look inside $r_{\rm h}$ to find the PDMF. 
Outside of $r_{\rm h}$ there is a paucity of massive stars and therefore 
a deficiency of young WDs, flattening the inferred IMF, and conversely we 
would expect an over-abundance of young WDs in the very central regions of 
the cluster, slightly offset by the presence of the majority of the very 
massive old WDs. 
Thus, assuming that our observations are accurate enough that we can remove a 
substantial fraction of the double-WDs from the LF, then using the LF 
of WDs residing in the $0.5 < r/r_{\rm h} < 1.0$ portion of a young cluster 
should provide a reliable age (although the non-standard single-WDs 
still remain -- see Section~\ref{s:clupar}). 
If double-WDs cannot be removed then binary evolution, as well as dynamical 
effects, must be accounted for in the fitting process. 
Factors such as the relative fraction of DA to DB WDs also play  
a role in determining the shape of the LF \citep{han02}. 
 
\subsection{Metallicity}
\label{s:metal}

The metallicity of the cluster population influences the LF fitting process 
in a number of ways. 
Firstly, the upper mass limit for a ZAMS star that will evolve to produce 
a WD has been shown to vary with $Z$: it decreases from $8.0 M_\odot$ for 
$Z = 0.02$ to $6.8 M_\odot$ for $Z = 0.0001$ in our adopted stellar 
evolution algorithm. 
This mass limit is found by inspecting detailed stellar models to determine 
at what mass carbon ignites in the core of an asymptotic giant branch (AGB) 
star, and the subsequent burning of this carbon produces an ONe  
core larger than the Chandrasekhar mass \citep{pol98}. 
The actual upper mass limit may be slightly less than this depending on the 
mass-loss rate assumed for stars on the AGB\footnote{ 
\citet{hur00} provides a description of the mass-loss rate used in the 
stellar evolution algorithm and also a more complete description of the 
WD upper mass limit.}. 
Secondly, the mass of a WD at birth depends on competition between the 
rate of growth of the degenerate core of its AGB star progenitor and the 
rate at which the envelope of the star is depleted by mass-loss 
-- both of which depend to some degree on $Z$. 
Finally, the lifetime of a MS star depends on its initial composition, 
shortening with decreasing metallicity for the range of initial masses 
that produce WDs. 
For example, the MS lifetime of a $2 M_\odot$ star decreases by $\sim 60\%$ 
when the metallicity of the star is reduced from $Z = 0.02$ to 0.0001. 
This has the effect of lowering the MS turn-off mass for low metallicity 
clusters and increasing the age of the WDs relative to MS stars. 
All of these effects are naturally accounted for in the LF fitting process 
provided that the correct metallicity is used when generating the 
theoretical LF. 
Of the four $N$-body simulations described by \citet{sha02}, each starting 
with $N = 20\,000$ and $f_{\rm b} = 10\%$, two were performed with 
$Z = 0.004$ while the other two had solar metallicity. 
We find that regardless of metallicity, the LFs for the standard single WDs 
at $4\,$Gyr in each of these simulations are best fit by an IMF slope of 
$\alpha \simeq 3.05$, noting that the range of ZAMS masses producing WDs at 
this time for $Z = 0.004$ is $1.2 - 7.0 M_\odot$. 
We also find that metallicity has a negligible effect on the WD mass 
fraction of a cluster (see next section), in agreement with \citet{hip98}.

\section{The Cluster White Dwarf Mass Fraction}
\label{s:wdmfr}

\begin{figure}
\plotone{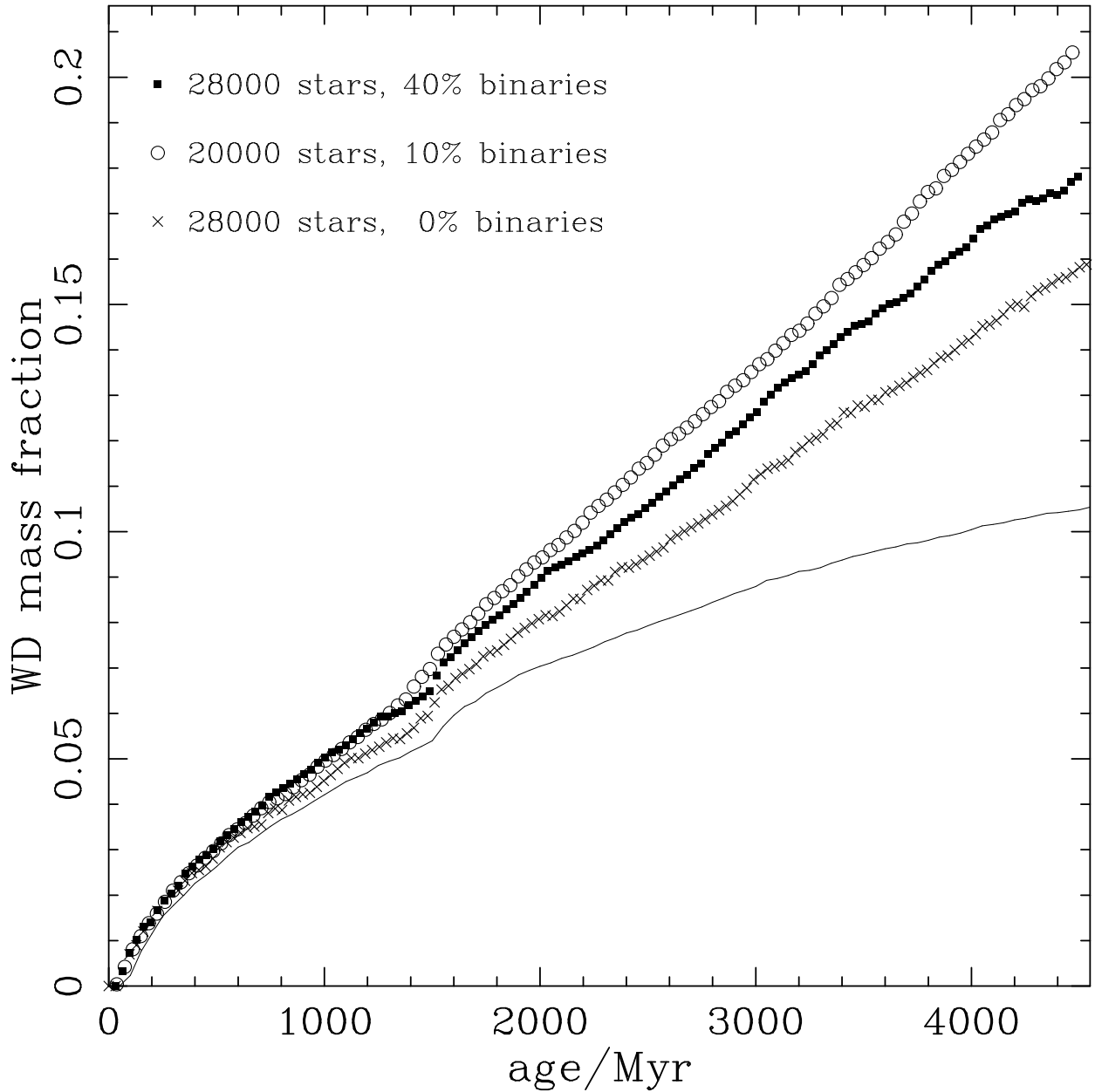}
\caption{
The WD mass fraction as a function of cluster age for simulations with 
$28\,000$ stars and a 40\% binary fraction (solid squares), 
$20\,000$ stars and a 10\% binary fraction (open circles), 
and $28\,000$ with no primordial binaries ($\times$ symbols). 
The corresponding mass fractions for the same populations evolved outside 
of the $N$-body code are also shown (solid line). 
Note that the mass fractions have been normalized so that the 
non-dynamical populations produce the same mass fraction of WDs 
(although the difference at any particular time is never more than a 
few percent). 
The remarkable deficit in WD mass fraction at 40\% binaries relative 
to 10\% binaries is discussed in Section~\ref{s:wdmfr}. 
\label{f:fig8}}
\end{figure}

Figure~\ref{f:fig8} shows the fraction of the cluster mass contained in WDs 
as a function of time for the $N$-body simulations with 
$N = 28\,000$ and $f_{\rm b} = 40\%$ discussed in this paper. 
Also shown is the WD mass fraction, $f_{\rm WD}$, for the same primordial 
populations evolved without dynamics. 
It is clear that as time progresses, and the simulated clusters become 
dynamically more evolved, that the cluster environment has a significant 
effect on the measured WD mass fraction. 
More correctly it is a combination of the cluster environment and the 
environment that the cluster resides in that is producing this effect, 
i.e. mass-segregation causes low-mass MS stars to move to the outer 
regions of the cluster so that these stars are preferentially stripped 
from the cluster relative to heavier objects such as WDs 
(see Section~\ref{s:mseg}). 
Therefore, even though properties of the WD population such as its total 
mass are relatively unaffected by the dynamical evolution of the cluster, 
the WD mass fraction is affected and all clusters will become rich in WDs 
in the late stages of their evolution (see also Vesperini \& Heggie 1997; 
Portegies Zwart et al. 2001). 

For young open clusters, less than $\sim 3 - 4$ relaxation times old, 
the results of our simulations suggest that it is safe to assume that the 
WD mass fraction has been little affected by the kinematic evolution 
of the cluster \citep{hip98}. 
However, for older open clusters and for globular clusters it would be 
incorrect to make this assumption. 
\citet{ves97} used $N$-body models with $N = 4\,096$ stars to investigate 
$f_{\rm WD}$ after $15\,$Gyr of evolution for globular clusters born 
with $N \sim 10^5$ stars. 
For a cluster orbiting within the Galactic potential at a distance of 
$4\,$kpc from the Galactic centre they found $f_{\rm WD} = 0.277$ 
(with some dependence on the IMF and density profile chosen for the model). 
Checking their results by performing additional simulations of $N = 8\,192$ 
and $N = 16\,384$ stars they found $f_{\rm WD} = 0.345$ and 
$f_{\rm WD} = 0.422$, respectively. 
This lack of convergence for increasing $N$ demonstrates the perils of scaling 
the results of small-$N$ simulations to large-$N$ clusters. 
However, owing to the difficulty in performing direct $N$-body models of 
globular clusters, scaling is often unavoidable and in that case our 
value of $f_{\rm WD} \simeq 0.18$ at $4\,$Gyr may be taken 
as a lower limit of the true value for an old globular cluster. 

\citet{ves97} showed that the position of a star cluster in the Galaxy 
is a factor in determining its WD mass fraction. 
They found that clusters orbiting at $4\,$kpc have $f_{\rm WD}$ increased 
by more than a factor of two over clusters at $16\,$kpc. 
Therefore, we would expect globular clusters such as M4 which has a 
perigalacticon of $1\,$kpc \citep{pet95}, to exhibit dynamically enhanced 
WD mass fractions. 
For M4 this is supported by observations of its MF for stars less massive than 
$0.65 M_\odot$ having a slope of only $\alpha = 0.75$ \citep{ric02}. 
In light of these results it would seem that the assumption by \citet{hip98} 
that the WD mass fraction of M4 has not been affected by the kinematic evolution 
of the cluster is in error. 
{\it We urge anyone utilising observations of the WD mass fraction in dynamically 
evolved star clusters, and especially those clusters orbiting near the 
Galactic centre, to account for the dynamical history of the cluster.} 
We note that the position in our simulated clusters where $f_{\rm WD}$ matches 
that of the non-dynamical population, after $4\,$Gyr, is at three half-mass 
radii from the cluster centre. 

After $4\,$Gyr of cluster evolution, or $\sim 12$ half-mass relaxation times, 
$f_{\rm WD}$ has approximately doubled in comparison to the same population 
evolved without dynamics. 
This translates to an error of $\Delta \alpha \simeq 0.2$, or 10\%, in 
the slope of the inferred power-law IMF if the dynamical evolution is 
ignored. 
\citet{por01} demonstrated that the dynamical evolution of a cluster has 
little effect on the numbers of giants and WDs present in the population 
and therefore that the relative number of these stars may be used to 
constrain the IMF of a star cluster. 
However, in making this point \citet{por01} were only drawing upon the results 
of small-$N$ simulations at $0.6\,$Gyr (after $\sim 6$ relaxation times). 
Our simulations with $N = 28\,000$ and $f_{\rm b} = 40\%$ 
also show minimal modification of the giant and WD numbers 
for the same degree of dynamical evolution ($\sim 2\,$Gyr for our larger $N$) 
but after $4\,$Gyr this is no longer true: only 48\% of the expected 
number of giants, and 50\% of the expected number of WDs, are present in the 
cluster at this time. 
Remarkably the number ratio of the two populations is still intact and 
would provide a reliable estimate of the IMF. 
Depletion of the number of WDs is primarily the result of WDs escaping 
from the cluster -- this accounts for 80\% of the decrease -- with the 
remainder of the depletion explained by enhanced 
binary interaction. 
Giants are much less likely to escape from the cluster because their lifetimes 
are extremely short compared to MS and WD lifetimes. 
As such, only 4\% of the expected population of giants at $4\,$Gyr was lost 
as a result of giants escaping the cluster. 
An additional 20\% of the expected giants were lost as a result of MS stars 
that would have evolved to appear as cluster giants at $4\,$Gyr escaping 
prior to this time. 
The remaining 23\%, i.e. $\sim 50\%$ of the decrease, is explained by the 
depletion of giants, or their MS progenitors, in 3- and 4-body encounters. 

In Figure~\ref{f:fig8} we also show the evolution of $f_{\rm WD}$ with time 
for simulations with $N = 20\,000$ and $f_{\rm b} = 10\%$, and for simulations 
with $N = 28\,000$ and no primordial binaries. 
We believe that the difference in $f_{\rm WD}$ between the two different 
families of $N$-body simulations with primordial binaries is real and 
can be explained in terms of binary fraction, and to a lesser extent 
the initial period distribution assumed for the binaries. 
Consider that a non-interacting, i.e. wide, binary composed of two $2.0 M_\odot$ 
stars will evolve to contribute two $\sim 0.65 M_\odot \,$WDs to the 
cluster WD population. 
If instead the binary components do interact and merge to form a single star 
(initial periods less than $\sim 2200\,$d) then only one WD with a mass in 
the range $\sim 0.8 - 0.9 M_\odot$, depending on the binary period, will 
be contributed. 
Therefore, the evolution of close binaries can decrease the mass in WDs. 
As a result the simulations with a large binary fraction (40\%) 
experience a larger relative decrease in the WD mass fraction owing to 
binary evolution than those with a moderate binary fraction (10\%). 
A larger binary fraction also provides more scope for the cluster to 
increase the fraction of close binaries through 3- and 4- body encounters. 

Comparisons between the simulations with and without primordial binaries 
simply using the cluster age can be misleading because the latter are 
dynamically less evolved at an age of $4\,$Gyr. 
Without binaries the half-mass relaxation timescale at $4\,$Gyr is 
$\sim 430\,$Myr which is twice as long as the timescale found for 
both families of simulations with primordial binaries. 
The primary reason for the difference is that clusters with a significant 
primordial binary population suffer a higher rate of mass loss from 
the cluster: 40\% of the initial cluster mass remains at $4\,$Gyr for 
the $f_{\rm b} = 0\%$ clusters, 27\% for $f_{\rm b} = 10\%$, and 
21\% for $f_{\rm b} = 40\%$. 
An enhanced rate of escaping stars via velocity kicks obtained in 
3-body interactions is the explanation for this trend. 
If we instead make a comparison of $f_{\rm WD}$ when the simulated clusters 
are all at the same dynamical age (10 half-mass relaxation timescales old) 
then we find 0.168 for $f_{\rm b} = 0\%$ ($4\,730\,$Myr), 0.170 for 
$f_{\rm b} = 10\%$ ($3\,720\,$Myr), and 
0.147 for $f_{\rm b} = 40\%$ ($3\,510\,$Myr).

\section{Discussion and Further Analysis}
\label{s:disc}

Using the results of $N$-body simulations to investigate and understand the 
nature of WD populations in star clusters, in terms of appearance in the CMD, 
the luminosity function, and mass fraction, appears to be a worthwhile approach. 
However, to this point we have concentrated on one particular type of simulation 
with its unique set of initial conditions. 
Variations to the model parameters, which range from the initial setup of the 
cluster to aspects of binary evolution, have the capability to influence the 
results. 
Here we discuss which parameters are of greatest importance and look at how 
our results can be utilised to interpret observations of WD populations in 
star clusters. 

\subsection{Variation of Cluster Parameters}
\label{s:clupar}

We have already seen in Section~\ref{s:wdmfr} that the primordial binary 
fraction of a star cluster plays a role in determining the evolutionary 
characteristics of the cluster, such as the escape rate of the stars and 
the relaxation timescale, as well as affecting aspects of the stellar 
populations, namely the WD mass fraction. 
Similarly we would expect the binary fraction to have an impact on the 
shape of the WD luminosity function and the appearance of the WD sequence 
in the CMD. 
In Tables~\ref{t:table2} and \ref{t:table3} we have replicated the luminosity 
function results of Table~\ref{t:table1} but this time for the simulations 
with $f_{\rm b} = 10\%$ and $f_{\rm b} = 0\%$, respectively. 
For $f_{\rm b} = 10\%$ the trends in the LF data are similar to those found 
for the simulations with $f_{\rm b} = 40\%$ but in general the behaviour is 
less extreme. 
As an example, the LF for standard single-WDs at $4\,$Gyr is best fit by 
$\alpha = 3.05$, as opposed to $\alpha = 3.15$ for $f_{\rm b} = 40\%$, so 
it is closer to the non-dynamical expected value. 
The probability of the fit being a good one is also higher which is an 
indication of a lower degree of contamination in the WD sequence. 
The reason for this is most likely a combination of the $f_{\rm b} = 10\%$ 
clusters having lost less of their stars at the same age and also that a 
smaller fraction of 
standard single-WDs have been affected by dynamical interactions. 
We note that the clusters with only 10\% primordial binaries have a smaller 
total number of stars than those with 40\% primordial binaries. 
This may also be a factor in any differences between the two types of simulation 
but on the other hand the total number of systems, single stars and binaries, 
is equal. 

For the simulations without primordial binaries the first binary formed 
after $\sim 300\,$Myr of evolution, which is roughly the half-mass 
relaxation timescale at that time. 
The number of binaries in the simulation jumps sharply at the time of 
core-collapse ($\sim 1\,400\,$Myr, or $\sim 3$ half-mass relaxation times) 
but even so, the number is only seven, or 0.03\%. 
The binary frequency subsequently reaches a peak of 0.07\% after 
$\sim 3\,500\,$Myr and 
basically stays at this value for the remainder of the simulation. 
Owing to the lack of binaries in these simulations it is not surprising to see
minimal contamination of the WD LF by double-WDs and non-standard single-WDs 
-- the LF results in Table~\ref{t:table3} for all WDs and standard single-WDs 
are identical. 
However, the shape of the WD LF has still been affected by the dynamical 
evolution of the cluster and the behaviour is similar to that found for 
the $f_{\rm b} = 10\%$ simulations. 

We note that for all of the LF fits described in 
Tables~\ref{t:table1}-\ref{t:table3} the 
value of the $\chi^2$ statistic for the fit is less than that given by 
the $\sqrt{N}$ uncertainty in the data points, except in two cases 
-- the $r > r_{\rm h}$ fits for all WDs and standard single-WDs at $1\,$Gyr 
for the $f_{\rm b} = 10\%$ simulations. 
Both LFs sufferred from a distinct lack of hot WDs for $M_V < 13$ which meant 
that only four data points were available to be fitted. 

In Section~\ref{s:obclue} we found that after $1$ and $2\,$Gyr of cluster 
evolution the best place to look for the true non-dynamical PDMF of the 
standard single-WDs was in the region $0.5 < r/r_{\rm h} < 1.0$, in the case 
of the $f_{\rm b} = 40\%$ simulations. 
This is also true for the other simulations that we have considered although 
for $f_{\rm b} = 0\%$ at $1\,$Gyr it is best to look closer to $0.5 r_{\rm h}$, 
and for both $f_{\rm b} = 0\%$ and $f_{\rm b} = 10\%$ clusters at $2\,$Gyr 
it is best to look closer to $r_{\rm h}$. 
For all the simulation types it is not possible to recover the PDMF after 
$4\,$Gyr of evolution. 

An interesting question is whether we can quantify our findings on the 
contamination of the WD sequence in such a way as to help observers of 
open clusters produce clean WD LFs? 
As shown in Table~\ref{t:table4} the level of contamination clearly 
increases with an increasing cluster primordial binary fraction, both in 
terms of the number of non-standard single-WDs and double-WDs produced, 
and the dynamical removal of standard single-WDs. 
We have also included in Table~\ref{t:table4} the ratio of double-WDs to 
single-WDs, and the ratio of non-standard single-WDs to all single-WDs, 
for the three distinct types of simulation. 
Unfortunately there is no clear relation between these two numbers, except 
that as one increases so does the other. 
One or two more data points and more simulations to decrease the noise in 
these results may lead to a more promising result. 
In the meantime the numbers presented in Table~\ref{t:table4} should prove 
useful, especially if observations are good enough to separate the 
majority of double-WDs from the single-WD sequence in the CMD and 
therefore gain an accurate estimate of the ratio of double-WDs to single-WDs. 
Alternatively the binary fraction of the cluster, or at least a lower limit, 
may be known from observations of the main sequence 
(Montgomery, Marschall \& Janes 1993, for example). 
Either way, the results of our simulations can then be used to estimate 
what fraction of non-standard single-WDs are present and remove these 
from the LF, although the behaviour of this fraction with magnitude is 
required for this approach to be of most use. 
Our recommendation for anyone wishing to derive information from the WD LF 
of a dynamically evolved star cluster is that they request data for 
simulations that best match the parameters of the observed cluster 
(age, binary fraction, etc.). 

Any aspect of the initial conditions chosen for a particular simulation that 
has the potential to affect the lifetime of the cluster or the amount of 
dynamical activity also has the potential to alter the makeup of the 
resultant stellar populations, and therefore create uncertainty in our results. 
In addition to binary fraction the parameters that immediately spring to 
mind are the number of stars, the density profile, 
and the external potential in which the cluster will orbit. 
Indeed, \citet{ves97} have shown that both the shape of the cluster mass-function 
and the WD mass fraction are significantly affected by the starting value of $N$ 
and the position of the cluster in the Galaxy, while the choice of initial 
density profile is less important. 
However, these results are only for models of single stars and it is the 
lack of large-scale simulations with substantial binary populations and a 
realistic treatment of binary evolution that makes it difficult to quantify 
the broader impact of these parameters on the stellar populations of clusters. 
This is precisely what we have started to address in the work presented here but 
considering that each of our $f_{\rm b} = 40\%$ simulations took a minimum 
of four weeks to perform a full parameter study will take time. 
To investigate the actual behaviour of the WD population in a globular cluster 
we need to push the particle number of our simulations to at least $10^5$ 
and recently \citet{bau02} have taken steps in this direction. 
With a single GRAPE-6 board and a primordial binary fraction of only a few 
percent it is estimated that a full scale simulation with $N = 10^5$ will 
take a minimum of four months to evolve to $10\,$Gyr.

\subsection{Variation of Stellar and Binary Evolution Parameters}
\label{s:binpar}

In the {\tt NBODY4} code all aspects of standard binary evolution, 
i.e. non-perturbed orbits,
are treated according to the prescription described in \citet{htp02}.
The problem with modelling binary evolution, whether it be by a prescription 
based approach or using a detailed evolution code \citep{nee01}, is that the 
outcome is extremely dependent on the input parameters to the model, which 
are themselves uncertain. 
Furthermore, we are not even sure of how to model some of the processes that 
arise. 
Take, for example, common-envelope evolution which is assumed to occur when 
mass transfer becomes unstable. 
In this case a detailed model of the process is still beyond us and those 
working in the field of binary population synthesis cannot even agree 
on a standard form for a simple model \citep{ibe93,nel01,htp02}, let alone 
parameters of the model such as the common-envelope efficiency parameter, 
${\alpha}_{\rm CE}$. 
The uncertainty that this creates is substantial because in population synthesis 
the common-envelope phase is crucial for the production of binaries such as 
cataclysmic variables and short-period double-WDs. 
For the $N$-body simulations described in this paper we have used 
${\alpha}_{\rm CE} = 3.0$ as this was shown by \citet{htp02} to give 
good agreement with observationally determined Galactic formation rates 
of various binary populations, although in most cases the observational 
tests were not particularly stringent. 
For example, \citet{htp02} showed that using ${\alpha}_{\rm CE} = 1.0$ does not 
predict enough short-period ($< 10\,$d) double-WDs, compared to local observations, 
whereas ${\alpha}_{\rm CE} = 3.0$ does. 
\citet{htp02} investigated the influence of a number of model parameters on the 
predicted Galactic formation rate of double-WDs. 
They found that the rate was not particularly sensitive to the choice of model 
for tidal evolution of the binary system, the metallicity of the population, 
or the initial eccentricity distribution, and changed only marginally with 
variation in ${\alpha}_{\rm CE}$ (within reasonable bounds of course). 
However, the predicted rate decreased by more than an order of magnitude when 
the component masses of each binary were drawn independently from the same IMF, 
as opposed to assuming a uniform distribution for the mass-ratios of the 
binaries, $n(q)$. 
These findings were in agreement with the extensive study of double-WD formation 
rates performed by \citet{zha98}, using a different binary evolution model. 
In addition \citet{zha98} considered the existence of a stellar wind, the velocity of 
this wind, enhancement of the stellar wind by the presence of a close companion, 
and the mass transfer efficiency of stable Roche-lobe overflow, as variable 
parameters and found that none of these had a substantial impact on the predicted 
rates.  

In Table~\ref{t:table5} we list population synthesis results relating to double-WDs 
for the parameters of the binary population that we feel have the potential to 
change the outcome of our $N$-body simulations: 
${\alpha}_{\rm CE}$, $n(q)$, and the initial distribution of the 
binary periods (or separations). 
The standard model (STD) assumes the parameters used in the $N$-body simulations, 
namely ${\alpha}_{\rm CE} = 3.0$, $n(q) = 1$, and the log-normal distribution 
of orbital separations given by \citet{egg89}.  
In turn we then try models with ${\alpha}_{\rm CE} = 1.0$ (CE1), binary component 
masses drawn independently from the IMF (IMF), and orbital separations chosen 
from a uniform distribution in the natural logarithm of the separation (SEP). 
Shown in Table~\ref{t:table5} are the fraction of binaries that are double-WDs 
and the value of $\alpha$ returned by fitting the WD LF (with corresponding 
probability of goodness-of-fit), after $4\,$Gyr. 
We also show the results of $\chi^2$ tests between the WD LF of each model 
and that of the standard model. 
Clearly the choice of $n(q)$ is the most crucial as the IMF model biases binaries 
to having small mass-ratios and leads to a greatly reduced number of double-WDs. 
The random pairing of binary component masses from the IMF is not well supported 
by observations \citep{egg89,duq91} however, \citet{kro95} has shown that this 
assumption may be valid if most stars are formed in embedded star clusters. 
Furthermore, \citet{tou91} has shown how selection effects make it difficult to 
determine $n(q)$ from observations. 
The distribution of orbital separations assumed in the SEP model is also not 
well supported by observations \citep{egg89,duq91} but once again the 
observed data is poorly constrained. 
It is comforting to see that the choice of ${\alpha}_{\rm CE}$ has only a small 
effect on the number of double-WDs in the model and that the appearance of the 
WD LF is unaffected, at least in the absence of dynamical interactions. 

\begin{figure}
\plotone{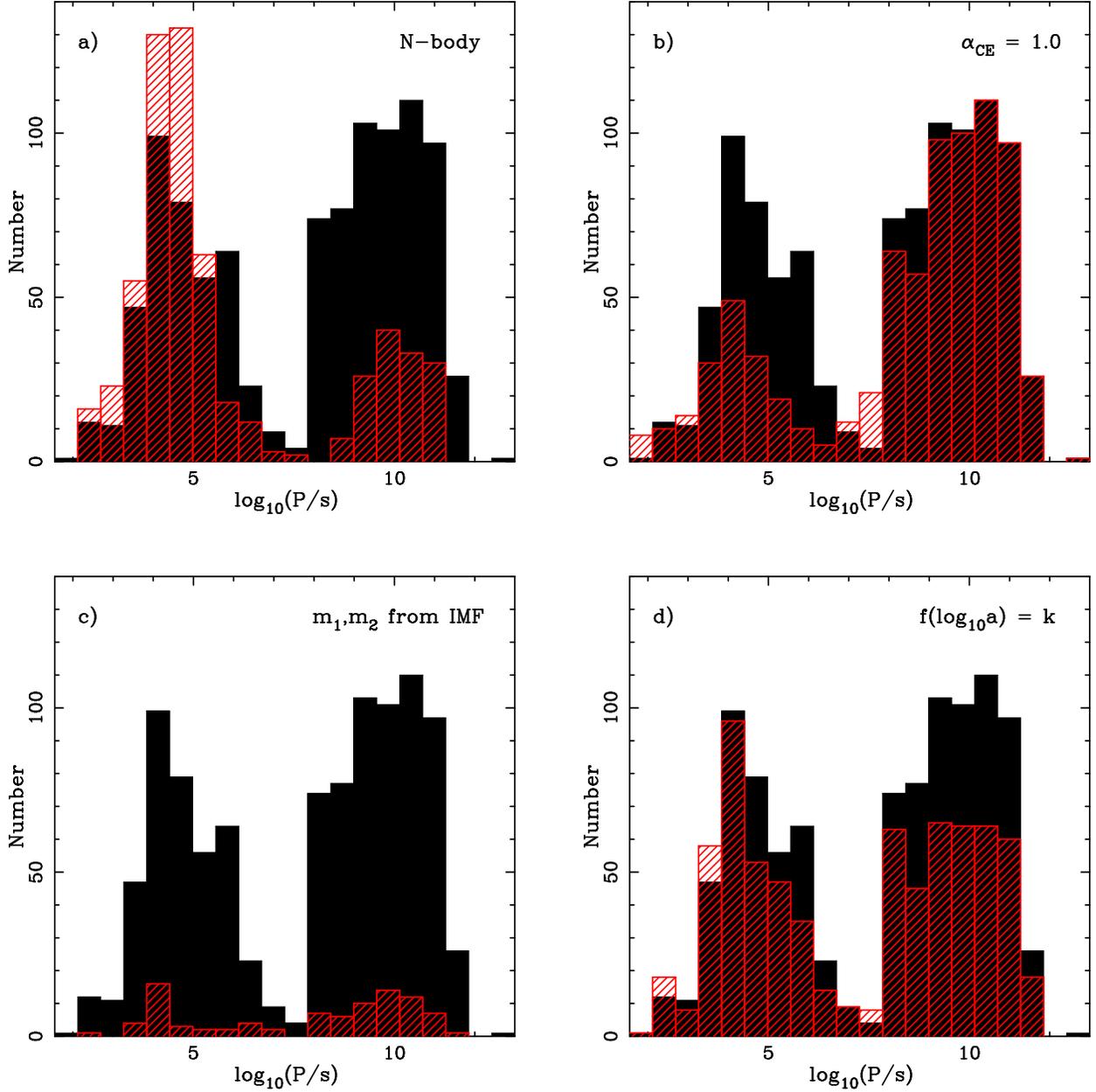}
\caption{
The distribution of orbital periods for double-WD populations at 
$4\,$Gyr of age. 
Panel (a) compares the distribution for the $N$-body simulations that 
started with $28\,000$ stars and a 40\% binary fraction (hatched) 
with the same population evolved outside of the cluster environment 
(solid, corresponds to population synthesis model STD). 
Panels (b), (c), and (d), compare the STD model (solid) with the 
CE1, IMF, and SEP models, respectively (hatched in all cases). 
\label{f:fig9}}
\end{figure}

When considering the implications of the various models for the results of our 
$N$-body study we must also look at the distribution of orbital periods for the 
double-WD populations of each model. 
This will help to determine how potential modification 
of the double-WD population via interaction with the cluster environment may be 
affected. 
Figure~\ref{f:fig9}a compares the distribution of double-WD periods at $4\,$Gyr 
arising from the STD model with that of the $N$-body simulations that started 
with $N = 28\,000$ and $f_{\rm b} = 40\%$. 
Thus the initial conditions of the two models are identical and the only 
difference is the presence of the cluster environment for the latter. 
For the $N$-body simulations it is clear that dynamical interactions between 
the binary population and other 
cluster stars (or binaries) is effective in destroying the wide double-WD 
population, as expected, and in enhancing the number of close double-WDs 
(as described by Shara \& Hurley 2002). 
In Figure~\ref{f:fig9}b we compare the STD and CE1 population synthesis models. 
The population of very wide double-WDs is unaffected by the change in 
${\alpha}_{\rm CE}$ as their evolution did not involve a common-envelope event. 
Importantly the relative number of double-WDs in the intermediate period range 
is reduced for ${\alpha}_{\rm CE} = 1.0$ and it is this population that is most 
likely to be modified by the cluster environment and lead to an enhancement of 
short-period WDs. 
Therefore, our results are sensitive to changes in ${\alpha}_{\rm CE}$, with a 
lower ${\alpha}_{\rm CE}$ leading to less contamination of the WD sequence by 
non-standard single-WDs. 
As expected, comparison of the STD and IMF models (see Figure~\ref{f:fig9}c) 
shows that the number of 
double-WDs is greatly reduced across the entire period range when the binary 
component masses are randomly assigned, reinforcing that adoption of this 
initial condition would lead to substantially less interesting results regarding 
double-WDs. 
The SEP model also shows a decrease in the number of intermediate-period 
double-WDs (see Figure~\ref{f:fig9}d), 
but not to the same degree as seen in the CE1 model. 
The decrease in the number of wide double-WDs is of little consequence for 
the $N$-body results. 

The bottom line is that uncertainty in the parameters of binary evolution and the 
initial conditions of the binary population leads to uncertainty in the 
results of $N$-body simulations relating to binary populations. 
However, to explore the parameter space of binary evolution within the framework 
of $N$-body simulations would be an inefficient use of computational resources. 
What we can do is to use the results of population synthesis calculations to 
determine which parameters are of greatest importance for future investigation. 
In the meantime we can also hope that observational constraints on certain 
parameters will improve, as will our understanding of processes such as 
common-envelope evolution. 
What is beyond question is that the cluster environment is very effective in 
modifying the evolution of the binaries it contains. 

The average WD mass of any stellar population depends to a large extent on the 
initial-final mass relation (IFMR) which links the ZAMS mass of a WD progenitor 
to the mass of the WD. 
The IFMR that results from the stellar evolution algorithm utilised in {\tt NBODY4} 
\citep{hur00} is biased towards higher WD masses, for ZAMS masses greater than 
$\sim 3 M_\odot$, 
than the semi-empirical IFMR derived by \citet{wei87} or the theoretical IFMR 
proposed by Han, Podsiadlowski \& Eggleton (1994, hereinafter HPE). 
We note that the IFMR produced by the stellar evolution algorithm is not 
pre-supposed: it is a natural consequence of the combined effects of the 
mass-loss prescription adopted for AGB stars and the evolution of the  
core-mass for AGB stars, as indicated by stellar models.  
It is also in good agreement with data on WDs observed in the young open 
cluster NGC$\,2516$ \citep{jef97}. 
After $4\,$Gyr of evolution for a non-dynamical population of single stars 
drawn from the \citet{kro93} IMF the average WD mass of the population is 
0.66 when using the SSE algorithm of \citet{hur00} and 0.62 if the HPE 
IFMR is adopted, i.e. a difference of 6\%. 
For the simulated clusters that started with $28\,000$ stars and 
$f_{\rm b} = 40\%$ we found that the average WD mass and the average mass 
of the single MS stars reached equality after $\sim 5\,050\,$Myr. 
If instead the HPE IFMR had been used then the crossover would have occurred 
$\sim 250\,$Myr earlier, i.e. a 5\% error. 
An area where a change in the IFMR has the potential to make a noticeable 
difference is the WD mass fraction of the cluster, $f_{\rm WD}$. 
However, for all of the $N$-body simulations we have presented the value of  
$f_{\rm WD}$ that we estimate if we had instead used the HPE IFMR in the 
models is in good agreement with the results of Section~\ref{s:wdmfr}: 
for cluster ages in excess of $400\,$Myr 
it is always within 5\% of the quoted values. 
So, all in all we do not expect a change in the IFMR to seriously alter 
our findings, especially at cluster ages where the corresponding MS 
turn-off mass is below $3 M_\odot$. 
One aspect that we have not considered in this analysis is that a decrease 
in the mass of a WD mass leads to an increase in its radius and therefore 
makes it more likely to interact with its companion, if indeed the WD is 
a member of a binary. 
In the SSE algorithm a $4 M_\odot$ ZAMS star would produce a 
$\sim 0.8 M_\odot$ WD, whereas use of the HPE IFMR would lead to a reduction 
of the WD mass by a factor of 7\%, corresponding to an increase of 5\% 
in the WD radius. 
To fully test the impact of this change on the results of our $N$-body 
simulations would require performing new simulations that adopted the 
HPE IFMR throughout. 
However, considering that agreement between the various IFMRs is quite 
good for ZAMS masses below $3 M_\odot$, or cluster ages greater than 
$\sim 500\,$Myr, we do not believe that this is a necessary course of 
action.

\subsection{Comparison with Observed Data Sets}
\label{s:obscmp}

In recent years increased interest in the WD populations of star clusters, 
due in part to their potential use as stellar chronometers, has lead to 
an increase in the available data relating to these populations 
(for recent reviews see von Hippel 1998; Koester 2002). 
The quality of these data sets has also improved but unfortunately not 
to the level where it allows rigorous comparison with our simulated 
clusters. 
Take, for example, the case of NGC$\,2420$ \citep{hip00} where the use 
of HST has allowed the detection of eight WDs and the calculation of an 
age for the cluster largely independent of stellar models. 
This number of WDs is far too small to enable a meaningful statistical 
comparison with either the WD CMD or the WD LF of our models. 
On the other hand, observations of M67 \citep{ric98} using the 
Canada-France-Hawaii Telescope have produced a much larger sample of WDs 
(of the order of a 100) but problems of background contamination and 
the statistical subtraction of background sources persist. 
As a result, the WD LF is only a statistical representation of the true 
WD LF of the cluster and it is not possible to identify the actual WD 
sequence in the CMD (as far as being sure that a particular point corresponds 
to a cluster member). 
Bearing in mind that the quality of the observed data sets will only 
improve in the near future we shall persist and attempt some example 
comparisons using the existing data 
in order to demonstrate what may be gleaned from such an approach. 
Necessarily we will focus on intermediate age open clusters. 

\citet{ric98} present a WD LF for M67 that after removal of background 
galaxies and field stars, and correction for incompleteness, contains 
85 WDs down to the termination point of the cooling sequence. 
For comparison with the LFs of our $N$-body models at $4\,$Gyr we have 
converted the observed LF to absolute magnitudes, using the distance modulus 
given by \citet{ric98}, and normalized the simulated and observed LFs so that 
the peaks match in terms of magnitude and number. 
This required a shift of $\sim 0.1\,$magnitudes which is representative 
of uncertainty in the cluster age: \citet{ric98} find a WD cooling age of 
$4.3\,$Gyr and a MS turn-off age of $4\,$Gyr. 
The normalization reveals that the observations are incomplete by a factor 
of $\sim 20$ in the next faintest magnitude bin after the peak and that 
the observed LF has too many hot WDs (by a factor of 4 for $M_V < 12.5$ bins), 
assuming that the simulated LFs are to be believed. 
Performing a $\chi^2$ test with each of our three distinct $N$-body data sets 
in turn shows that the observed LF is most likely drawn from the same 
population as the models with a 40\% binary fraction (probability of 0.46) 
-- a heartening result considering that \citet{ric98} quote a binary fraction 
of 50\% for M67. 
Based on the number of giants present in the M67 CMD \citet{ric98} estimate 
that they have only found $40\% \pm 10\%$ of the expected number of WDs 
with cooling ages $\le 1\,$Gyr. 
They also find that the mass fraction of WDs is 0.09 which is approximately 
half the number predicted by our 40\% binary model at $4\,$Gyr, which itself 
is very similar to the observed parameters of M67. 
Therefore, we agree that M67 appears deficient in hot WDs, indicating that 
dynamical interactions have been efficient in destroying hot WDs in this 
cluster. 

The problems with foreground and background contamination that plague 
the interpretation of observations of open clusters such as M67 are 
not as severe in the case of the rich open cluster NGC$\,6819$ 
\citep{kal01b}. 
Even though the photometry for NGC$\,6819$ does not reach down to the 
termination of the WD sequence, the improved quality of the data 
(and better choice of filters) 
makes it possible to compare the upper part of the WD sequence with our 
cluster models. 
Figure~\ref{f:fig10}a shows the potential WD candidates found in the 
CMD of NGC$\,6819$ by \citet{kal01b}, including only objects that have 
stellarity index of 0.75 or greater (where 0 is most likely a galaxy 
and 1 is definitely a star). 
We note the existence of what appears to be two distinct WD sequences, 
where the brighter sequence is not consistent with being a population 
of background WDs with higher reddening nor can it be reproduced by a 
spread in WD masses (J. Kalirai, private communication, 2002).  
The age of NGC$\,6819$ is reported to be $\sim 2.5\,$Gyr \citep{ros98,kal01b} 
so in Figures~\ref{f:fig10}b, c, and d, we show the WD CMDs for our 
simulated clusters with 0\%, 10\%, and 50\% binaries, respectively, at an 
age of $2.5\,$Gyr.  
For the sake of comparison we have added photometric error to the 
$N$-body data points in order to 
give an impression of how an observed WD sequence would appear 
(for a large open cluster) if all WDs were detected. 
The photometric error has been estimated using a rather simple expression 
that is basically Gaussian noise with a spread of $\pm A \sigma$ mags, 
where we have assumed $\sigma = 0.05$ in this instance. 
The value of $A$ is given by $(m_x/10)^4$, where $m_x$ is the magnitude of 
interest, and it is used to mimic the increasing error for decreasing 
brightness. 
A simple comparison by eye of the CMDs shown in Figure~\ref{f:fig10} 
indicates that the bright sequence of WDs in the NGC$\,6819$ CMD 
could very well be the detection of a population of double-WDs in the 
cluster and, if this is the case, that NGC$\,6819$ most likely had 
a large fraction of primordial binaries. 
To aid the comparison we have merged Figures~\ref{f:fig10}a and \ref{f:fig10}d 
with the result shown in Figure~\ref{f:fig11}. 

Although these two short examples have not lead to any statistically significant 
conclusions we maintain that they have served the purpose of demonstrating that 
comparisons of $N$-body models with observations of star clusters will 
lead to a better understanding of the nature of cluster WD populations. 
For open clusters we now have the tools in place to produce realistic 
model clusters which can be compared with real clusters on a star-to-star 
basis. 
It will be particularly informative to tailor a set of simulations to an 
actual open cluster for which good quality data exists, such as NGC$\,6819$, 
with the initial conditions of the model hopefully constrained by 
observations. 
Such an approach will rely on close collaboration between stellar dynamicists 
and observers, and efforts in this direction are already underway. 

Finally we would like to comment on what our results can say about the 
excellent data set provided by HST observations of the globular cluster 
M4 \citep{ric02,han02} -- even though our $N$-body models fall far short 
of the particle number required for a globular cluster simulation. 
For M4 \citet{han02} used the method of fitting the entire WD LF to derive an 
age of $12.7 \pm 0.7\,$Gyr for the cluster. 
In doing so they assumed an IMF slope of $\alpha = 1.05$ and they claim that 
the result is robust to changes in $\alpha$ 
as long as it remains in the range of $0.7 - 1.1$. 
Their choice of $\alpha$ was based on a comparison
of the number counts of WDs to low-mass MS stars in the cluster \citep{ric02}. 
Our simulations have shown that as a star cluster evolves the value of the IMF 
slope inferred from the WD population steepens while the mass function of MS 
stars is flattened, primarily owing to stripping of stars from the cluster by 
the Galactic tidal field. 
M4 is not known to be a post-core-collapse cluster \citep{har96} so it is not 
dynamically old, but its current half-mass relaxation timescale is estimated 
to be of the order of $500\,$Myr \citep{har96}. 
A conservative estimate of its dynamical age is probably in the range of 
$4-5$ half-mass relaxation times old, considering that our $N$-body models 
have generally passed through core-collapse by this point, and as such our 
results would predict that the WD LF should be well represented by a value 
of $\alpha$ only slightly greater than the IMF value. 
However, considering the orbit of M4 and the findings of \citet{ves97}, we 
would expect the mass function of MS stars to have been significantly 
flattened at the low-mass end. 
Furthermore, \citet{kro93} have shown that for field stars the IMF slope 
for low-mass stars is considerably flatter 
($\alpha = 1.3$ for $M < 0.5 M_\odot$) than for stars in the WD-producing 
mass range. 
As a result we find it surprising that the WD LF of M4 could be well fitted 
using such a low value of $\alpha$. 
This could indicate that for M4 stellar interactions have been very effective 
in removing many of the hotter WDs (cf. the discussion of M67 above), an 
explanation that can also support arguments that the IMF of M4 was steeper 
than the PDMF quoted by \citet{ric02} while preserving the observed 
number counts. 
However, M4 is not a particularly dense globular cluster. 
Perhaps this is evidence that the IMF of M4 is different from that of field 
stars but it is more likely that subtle effects, such as the stellar evolution 
age of M4 being much greater than our $N$-body models, lead to differences 
in the evolution of the WD LF compared to our open cluster models.

\section{Summary}
\label{s:concl}

To first order the WD population of a star cluster is relatively 
unaffected by the dynamical evolution of the cluster. 
However, segregation of the progenitor stars of the WDs, or binaries 
containing these progenitors, towards the centre of the cluster 
and the stripping of low-mass stars by the Galactic tidal field, 
affects both the appearance of the WD population and the WD mass fraction. 
In this work we have utilised the results of realistic $N$-body simulations 
of large open clusters to illustrate and quantify the behaviour of 
WDs in dense star clusters. 

We find that the presence of a substantial binary population in a star 
cluster, and the interaction of this population with the cluster 
environment, leads to a noticeable contamination of the observed WD 
sequence in the CMD by non-standard single-WDs and double-WD binaries. 
Scatter in the WD sequence produced by this contamination makes it 
difficult to judge the true bottom of the standard WD sequence, 
especially if a large population of old double-WDs exists. 
This can lead to significant errors in the derived WD cooling age for 
the cluster. 
However, provided that the full extent of the WD sequence has been 
observed, the presence of non-standard single-WDs and double-WDs in 
the WD LF does not affect the location of the LF peak. 
So ages measured by this technique do {\it not} suffer any additional 
error. 
The shape of the LF {\it is} affected by the presence of non-standard WDs 
which produces an uncertainty in the inferred slope of the IMF. 
The amount of contamination of the WD LF by non-standard single-WDs is 
proportional to the number of double-WDs in the cluster which itself 
is linked to the primordial binary fraction of the cluster 
-- if the number of double-WDs observed in a real WD sequence is low 
then it is safe to assume that the number of non-standard single-WDs is 
also low. 
Furthermore, the dynamical evolution of the cluster hampers the use 
of WDs as tracers of the IMF. 

The results of our $N$-body simulations suggest that observations 
of star clusters should be conducted slightly interior to the half-mass 
radius of the cluster in order to best obtain information about the 
IMF from the WD LF. 
This region provides a compromise between our suggestion in 
Section~\ref{s:wdseq} that contamination of the WD sequence by 
double-WDs is less of a problem exterior to $r_{\rm h}$ and 
our findings in Section~\ref{s:lumfun} that the true present-day mass function 
for single-WDs is recovered for $0.5 < r/r_{\rm h} < 1.0$, 
at least for clusters less than $\sim 6$ half-mass relaxation times old. 
For dynamically evolved star clusters the WD LF cannot be used to 
recover information about the slope of the IMF, regardless of the 
primordial binary fraction of the cluster. 
Our findings are particularly instructive for future HST observations of 
globular 
clusters where it is only possible to observe a portion of the cluster. 
\citet{ves97} have shown that for clusters having undergone substantial 
dynamical evolution the PDMF for all cluster stars bears no relation to 
the IMF, even near the half-mass radius. 
This is especially true for globular clusters with orbits closer than
$8\,$kpc to the Galactic centre. 
Considering that large globular clusters are at an intermediate dynamical 
age, and that the PDMF of MS stars loses memory of the IMF earlier than 
does the WD population, then observations of the WD LF near the half-mass 
radius may be the best way to learn about the IMF of these objects. 
For open clusters it is generally possible to observe the entire 
cluster, at least for nearby clusters. 
In this case there is the option to restrict the LF to only include WDs 
found in a certain region of the cluster. 
However, considering that open clusters 
contain fewer stars than globulars, this is a counterintuitive action in 
terms of obtaining a statistically significant result. 
Furthermore, open clusters are observed to have relatively high binary 
fractions (Richer et al. 1998, for example) so contamination of the WD 
sequence by non-standard single WDs and double-WDs will be more of a problem. 
Therefore, we recommend that fitting of the WD LF for open clusters takes 
into account the evolution of binaries and the cluster environment.  

Our simulations have also shown that the WD mass fraction of a cluster is 
altered by the kinematic evolution of the cluster provided that it is 
more than a few relaxation times old. 
For a large open cluster after $4\,$Gyr of evolution, such as M67, the 
WD mass fraction is double the value expected from the same population 
evolved outside of the cluster environment. 
This enhancement of the WD mass fraction is primarily explained by the 
preferential escape of low-mass stars from the cluster and this 
result should be taken as a lower limit to the enhancement expected 
in moderate-size globular clusters. 
We find that enhancement of the WD mass fraction is not particularly 
sensitive to the primordial binary fraction of a star cluster. 
The expected number ratio of giants to WDs is preserved in our simulated 
clusters even though the two populations are affected in different ways 
by their residence in a cluster. 
Observation of this quantity is therefore one way to extract information 
about the IMF of an open cluster. 
Here, and in all instances, we must be very careful about scaling the results 
of our simulations to large-$N$ globular clusters. 
Additional depletion of giants may be expected in the centres of 
globular clusters which typically have much higher stellar densities than 
our models. 
Direct simulations of this type will be performed in the near future, 
providing additional information about the behaviour of WDs, and all 
populations for that matter, in star clusters. 
Future models will also address the potential uncertainty in our results 
eminating from uncertainties in the parameters of binary evolution.

\acknowledgments

\begin{figure}
\plotone{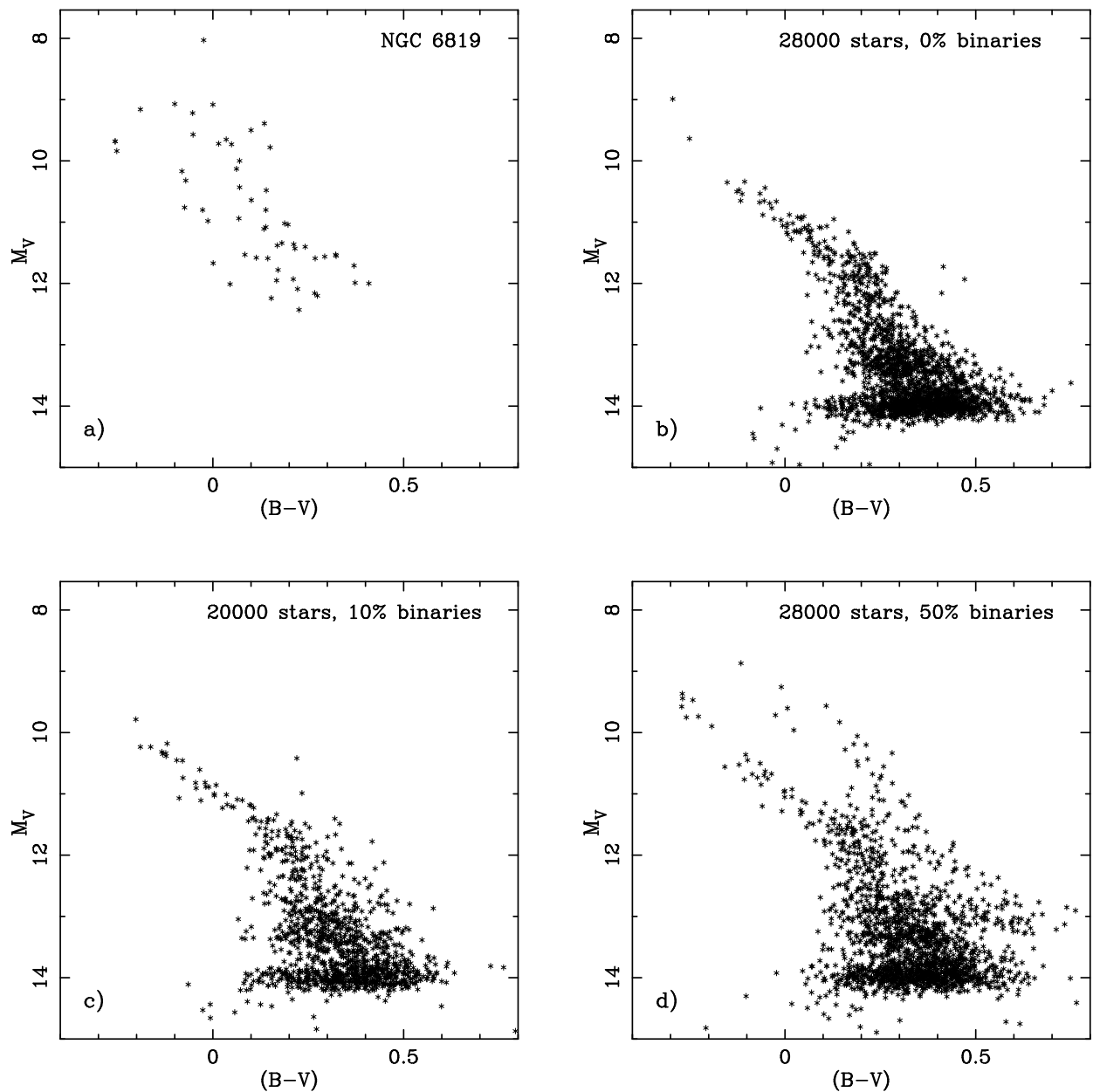}
\caption{
Cluster CMDs showing the WD sequence. 
Panel (a) shows the observed data points for NGC$\,6819$ after  
discarding objects with a stellarity index less 
than 0.75 \citep{kal01b}. 
Panels~(b), (c), and (d) show the $N$-body models with 0\%, 10\%, 
and 40\%, primordial binaries, respectively, at an age of 
$2.5\,$Gyr. 
The $N$-body data has simulated photometric error added. 
\label{f:fig10}}
\end{figure}

\begin{figure}
\plotone{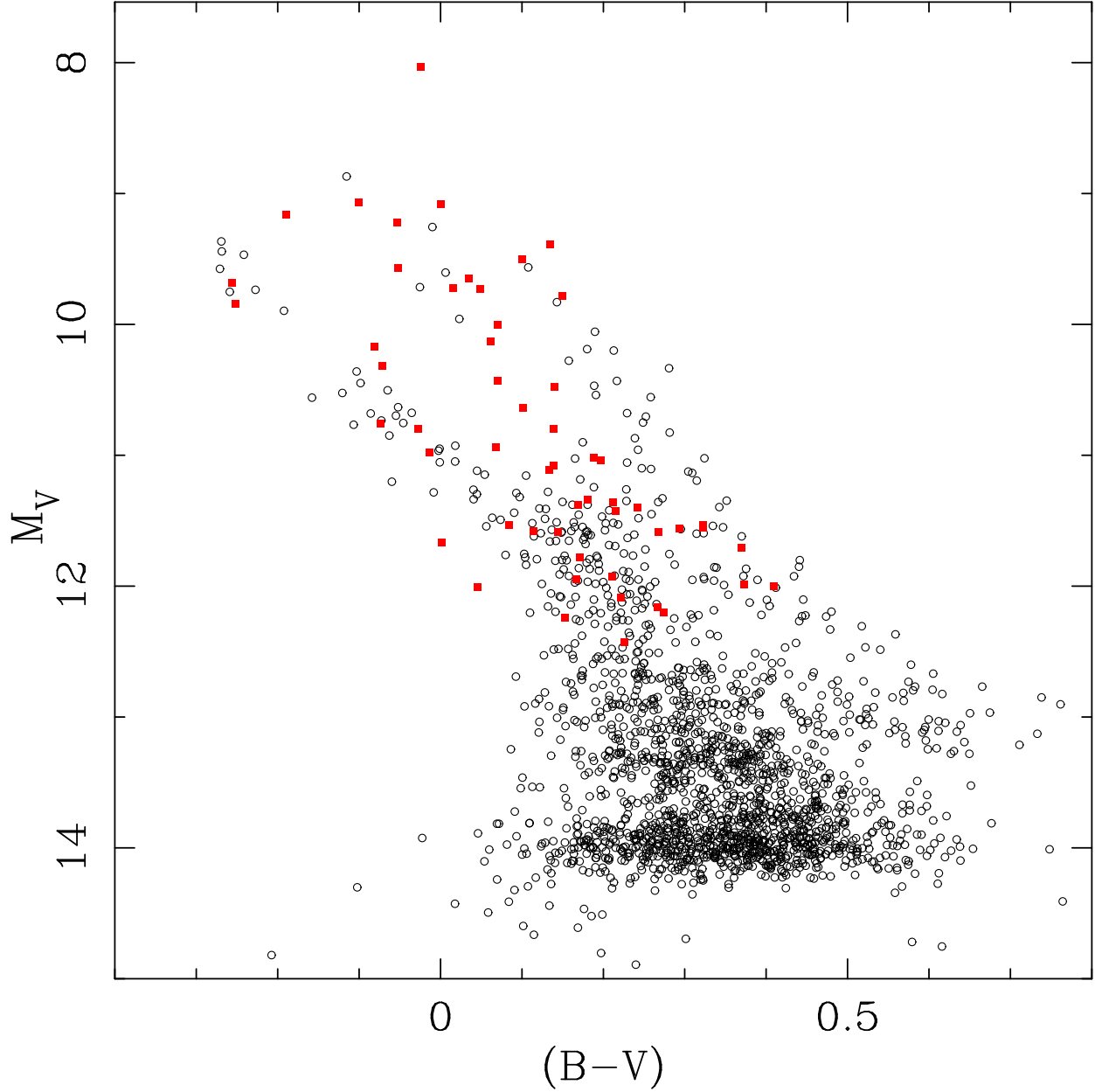}
\caption{
Same as Figure~\ref{f:fig10} but now with the NGC$\,6819$ 
data (solid squares) combined with the WD sequence from the 
$N$-body models that started with 40\% primordial binaries 
(open circles). 
\label{f:fig11}}
\end{figure}

We acknowledge the generous support of the Cordelia Corporation and 
that of Edward Norton which 
has enabled AMNH to purchase new GRAPE-6 boards and supporting hardware. 
We thank Brad Hansen for valuable advice on the cooling of white dwarfs 
and David Zurek for many helpful comments and for 
his help in developing Figure~\ref{f:fig10}. 
We also thank Jason Kalirai for making available his data on 
cluster white dwarfs and taking the time to help us understand the 
current status of the observations.  
This work was partially supported by NASA through Hubble Fellowship 
grant HST-HF-01149.01-A awarded to JRH by the Space Telescope Science 
Institute, which is operated by the Association of Universities for 
Research in Astronomy, Inc., for NASA, under contract NAS 5-26555.

\clearpage

\clearpage

\begin{deluxetable}{ccccccccccccc}
\tablecolumns{13}
\tablewidth{0pc}
\tablecaption{
Luminosity function data for simulations with 
$N = 28\,000$ and $f_{\rm b} = 40\%$. 
The results of the fitting process are given for LFs containing 
all WDs, including double-WDs, and those containing only the WDs 
identified as being standard single. 
We consider LFs for the entire cluster, and also for WDs interior 
or exterior to the cluster half-mass radius. 
In each case we give the best fitting $\alpha$ from a single power-law 
\citet{sal55} IMF, based on the minimum $\chi^2$, and the probability 
that this provides a good fit to the data. 
\label{t:table1}
}
\tabletypesize\footnotesize
\tablehead{
Age & \multicolumn{2}{c}{all WDs} &  
\multicolumn{2}{c}{($r < r_{\rm h}$)} & 
\multicolumn{2}{c}{($r > r_{\rm h}$)} & 
\multicolumn{2}{c}{st. sgl-WDs} & 
\multicolumn{2}{c}{($r < r_{\rm h}$)} & 
\multicolumn{2}{c}{($r > r_{\rm h}$)} \\ 
(Gyr) & $\alpha$ & Prob & $\alpha$ & Prob  & $\alpha$ & Prob &  
$\alpha$ & Prob & $\alpha$ & Prob  & $\alpha$ & Prob 
}
\startdata
1.0 & 2.75 & 0.93 & 3.55 & 0.24 & 2.40 & 0.28 & 2.70 & 0.87 & 3.15 & 0.89 & 
 2.15 & 0.31 \\
2.0 & 3.00 & 0.43 & 3.65 & 0.16 & 2.75 & 0.65 & 2.80 & 0.76 & 3.20 & 0.85 & 
 2.65 & 0.97 \\ 
4.0 & 3.75 & 0.30 & 3.90 & 0.86 & 3.55 & 0.26 & 3.15 & 0.65 & 3.30 & 0.94 & 
 3.00 & 0.46 \\ 
\enddata
\end{deluxetable}

\begin{deluxetable}{ccccccccccccc}
\tablecolumns{13}
\tablewidth{0pc}
\tablecaption{
Luminosity function data for simulations with 
$N = 20\,000$ and $f_{\rm b} = 10\%$. 
\label{t:table2}
}
\tabletypesize\footnotesize
\tablehead{
Age & \multicolumn{2}{c}{all WDs} &  
\multicolumn{2}{c}{($r < r_{\rm h}$)} & 
\multicolumn{2}{c}{($r > r_{\rm h}$)} & 
\multicolumn{2}{c}{st. sgl-WDs} & 
\multicolumn{2}{c}{($r < r_{\rm h}$)} & 
\multicolumn{2}{c}{($r > r_{\rm h}$)} \\ 
(Gyr) & $\alpha$ & Prob & $\alpha$ & Prob  & $\alpha$ & Prob &  
$\alpha$ & Prob & $\alpha$ & Prob  & $\alpha$ & Prob 
}
\startdata
1.0 & 2.70 & 0.92 & 3.30 & 0.69 & 2.00 & 0.10 & 2.70 & 0.88 & 2.90 & 0.89 & 
 2.00 & 0.08 \\
2.0 & 3.20 & 0.35 & 3.45 & 0.07 & 2.35 & 0.40 & 2.75 & 0.26 & 3.20 & 0.06 & 
 2.25 & 0.68 \\ 
4.0 & 3.40 & 0.62 & 3.55 & 0.77 & 3.25 & 0.83 & 3.05 & 0.82 & 3.30 & 0.93 & 
 3.00 & 0.93 \\ 
\enddata
\end{deluxetable}

\begin{deluxetable}{ccccccccccccc}
\tablecolumns{13}
\tablewidth{0pc}
\tablecaption{
Luminosity function data for simulations with 
$N = 28\,000$ and $f_{\rm b} = 0\%$. 
\label{t:table3}
}
\tabletypesize\footnotesize
\tablehead{
Age & \multicolumn{2}{c}{all WDs} &  
\multicolumn{2}{c}{($r < r_{\rm h}$)} & 
\multicolumn{2}{c}{($r > r_{\rm h}$)} & 
\multicolumn{2}{c}{st. sgl-WDs} & 
\multicolumn{2}{c}{($r < r_{\rm h}$)} & 
\multicolumn{2}{c}{($r > r_{\rm h}$)} \\ 
(Gyr) & $\alpha$ & Prob & $\alpha$ & Prob  & $\alpha$ & Prob &  
$\alpha$ & Prob & $\alpha$ & Prob  & $\alpha$ & Prob 
}
\startdata
1.0 & 2.70 & 0.97 & 2.95 & 0.88 & 2.15 & 0.47 & 2.70 & 0.97 & 2.95 & 0.88 & 
 2.15 & 0.47 \\
2.0 & 2.80 & 0.89 & 2.95 & 0.83 & 2.05 & 0.99 & 2.80 & 0.90 & 2.95 & 0.82 & 
 2.05 & 0.99 \\ 
4.0 & 3.05 & 0.66 & 3.15 & 0.90 & 3.05 & 0.80 & 3.05 & 0.67 & 3.15 & 0.91 & 
 3.05 & 0.79 \\ 
\enddata
\end{deluxetable}

\begin{deluxetable}{crrrccc}
\tablecolumns{7}
\tablewidth{0pc}
\tablecaption{
Contamination of the WD sequence at $4\,$Gyr for the various simulation 
classes. 
The first column describes the type of simulation. 
Columns~2 and 3 give the number of non-standard single WDs ($n_{\rm nss}$) 
and the number of double-WDs ($n_{\rm\scriptscriptstyle DWD}$) in the 
WD sequence. 
Column~4 shows the number of expected standard single-WDs ($n_{\rm ss}$) that 
are exchanged into binary systems and Column~5 shows the fraction of all 
exchange interactions that this comprises. 
Note that the numbers in Columns~2-4 are per 1000 standard single-WDs. 
Columns~6 and 7 show the ratio of non-standard single-WDs and double-WDs, 
respectively, to the total number of single-WDs 
($n_{\rm s} = n_{\rm ss} + n_{\rm nss}$). 
\label{t:table4}
}
\tabletypesize\footnotesize
\tablehead{
Simulation & $n_{\rm nss}$ & $n_{\rm\scriptscriptstyle DWD}$ & $n_{\rm ss,ex}$ & 
$n_{\rm ss,ex} / n_{\rm ex}$ & $n_{\rm\scriptscriptstyle DWD} / n_{\rm s}$ & 
$n_{\rm nss} / n_{\rm s}$ 
}
\startdata
$N = 28\,000$, $f_{\rm b} = 40\%$ & 694 & 230 & 162 & 0.49 & 0.136 & 0.410 \\ 
$N = 20\,000$, $f_{\rm b} = 10\%$ & 213 &  36 & 109 & 0.61 & 0.030 & 0.176 \\ 
$N = 28\,000$, $f_{\rm b} = \,\,\, 0\%$ &   8 &   4 &  42 & 0.62 & 0.004 & 0.008 \\ 
\enddata
\end{deluxetable}

\begin{deluxetable}{ccccc}
\tablecolumns{5}
\tablewidth{0pc}
\tablecaption{
Results for population synthesis models at an age of $4\,$Gyr. 
Column~1 gives the model name (see text for details) and Column~2 
gives the fraction of double-WD binaries present in the model.  
Column~3 shows the best fitting $\alpha$ from a single power-law 
\citet{sal55} IMF, based on the minimum $\chi^2$, to the WD LF 
and the probability that this provides a good fit to the data 
is shown in Column~4. 
The probablity that the WD LF of the model and that of the STD model 
are drawn from the same distribution is given in Column~5. 
\label{t:table5}
}
\tabletypesize\footnotesize
\tablehead{
Model & $f_{\rm\scriptscriptstyle DWD}$ & $\alpha$ & Prob & Prob 
}
\startdata
STD & 0.09 & 3.15 & 0.48 & - \\ 
CE1 & 0.07 & 3.15 & 0.75 & 0.98 \\ 
IMF & 0.01 & 2.55 & 0.99 & 0.00 \\ 
SEP & 0.06 & 2.90 & 0.88 & 0.09 \\ 
\enddata
\end{deluxetable}

\end{document}